\documentclass[10pt,conference]{IEEEtran}
\usepackage{cite}
\usepackage{amsmath,amssymb,amsfonts}
\usepackage{algorithmic}
\usepackage{graphicx}
\usepackage{textcomp}
\usepackage{comment}
\usepackage{xcolor}
\usepackage[hyphens]{url}
\usepackage{fancyhdr}
\usepackage{hyperref}

\title{A comprehensive study on ILP acceleration accounting for sparsity, area, energy, data movement using near-memory architecture}

\author{
  \ifdefined\hpcacameraready
    \IEEEauthorblockN{\hpcaauthors{}}
      \IEEEauthorblockA{
        \hpcaaffiliation{} \\
        \hpcaemail{}
      }
  \else
    \IEEEauthorblockN{\normalsize{Siddhartha Raman Sundara Raman, Lizy John, and Jaydeep P. Kulkarni} \\
      \IEEEauthorblockA{
      The University of Texas at Austin \\
      s.siddhartharaman@utexas.edu \\ This paper is an extended version of the paper published in IEEE HPCA 2025
      }
    }
  \fi 
}

\fancypagestyle{camerareadyfirstpage}{%
  \fancyhead{}
  
  \fancyhead[C]{
    \ifdefined\aeopen
    \parbox[][12mm][t]{13.5cm}{\hpcayear{} IEEE International Symposium on High-Performance Computer Architecture (HPCA)}    
    \else
      \ifdefined\aereviewed
      \parbox[][12mm][t]{13.5cm}{\hpcayear{} IEEE International Symposium on High-Performance Computer Architecture (HPCA)}
      \else
      \ifdefined\aereproduced
      \parbox[][12mm][t]{13.5cm}{\hpcayear{} IEEE International Symposium on High-Performance Computer Architecture (HPCA)}
      \else
      \parbox[][0mm][t]{13.5cm}{\hpcayear{} IEEE International Symposium on High-Performance Computer Architecture (HPCA)}
    \fi 
    \fi 
    \fi 
    \ifdefined\aeopen 
      \includegraphics[width=12mm,height=12mm]{ae-badges/open-research-objects.pdf}
    \fi 
    \ifdefined\aereviewed
      \includegraphics[width=12mm,height=12mm]{ae-badges/research-objects-reviewed.pdf}
    \fi 
    \ifdefined\aereproduced
      \includegraphics[width=12mm,height=12mm]{ae-badges/results-reproduced.pdf}
    \fi
  }
  \fancyfoot[C]{This paper}
}
\begin{document}
\maketitle

\ifdefined\hpcacameraready 
  \thispagestyle{camerareadyfirstpage}
  \pagestyle{empty}
\else
  \thispagestyle{plain}
  \pagestyle{plain}
\fi

\ifdefined\eaopen
\renewcommand{\hpcaheight}{12mm}
\fi


\begin{abstract}

 Integer Linear Programming (ILP) is widely used for solving real-world optimization problems, including network routing, map routing, and traffic scheduling. However, ILP algorithms are sparse and branch-intensive, making them inefficient on conventional CPUs and GPUs. Prior work has shown that large-scale ILP problems can require tens of hours of execution time even on massively parallel systems, limiting their applicability to time-sensitive decision-making workloads.

Existing ILP solvers such as Gurobi employ software-level optimizations to handle sparsity on CPUs, but still face throughput limitations. GPU-based ILP solvers are also constrained because GPUs are not well suited for sparse and branch-heavy workloads, leading to thread divergence, under-utilization of streaming multiprocessors, and frequent host-device interactions.

This paper presents SPARK, a sparsity-aware, reuse-aware, energy-efficient, reconfigurable near-cache ILP accelerator. SPARK repurposes the existing L1 cache in CPUs to provide near-cache acceleration with minimal hardware overhead of approximately 1.4\% of the CPU area. The architecture performs near-cache sparsity detection and sparsity-aware computation to reduce insignificant computations and data movement energy. SPARK also exploits computational reuse patterns in ILP algorithms to improve parallelism and efficiency. The proposed design supports both sparse and dense ILPs as well as Linear Programs (LPs).

Evaluations on real-world workloads from MIPLIB 2017 show that SPARK achieves up to 15× and 20× performance improvement, and up to 152× and 740× energy reduction compared to AMD Zen3 CPUs and NVIDIA Tesla V100 GPUs, respectively, for sparse ILPs. For sparse LPs, SPARK achieves 7–17× performance improvement and 103–250× energy reduction over CPU and GPU baselines, demonstrating the broad applicability of the proposed architecture.

\end{abstract}

\section{Introduction}

\begin{figure}[t]
\centering
\includegraphics[width=\linewidth]{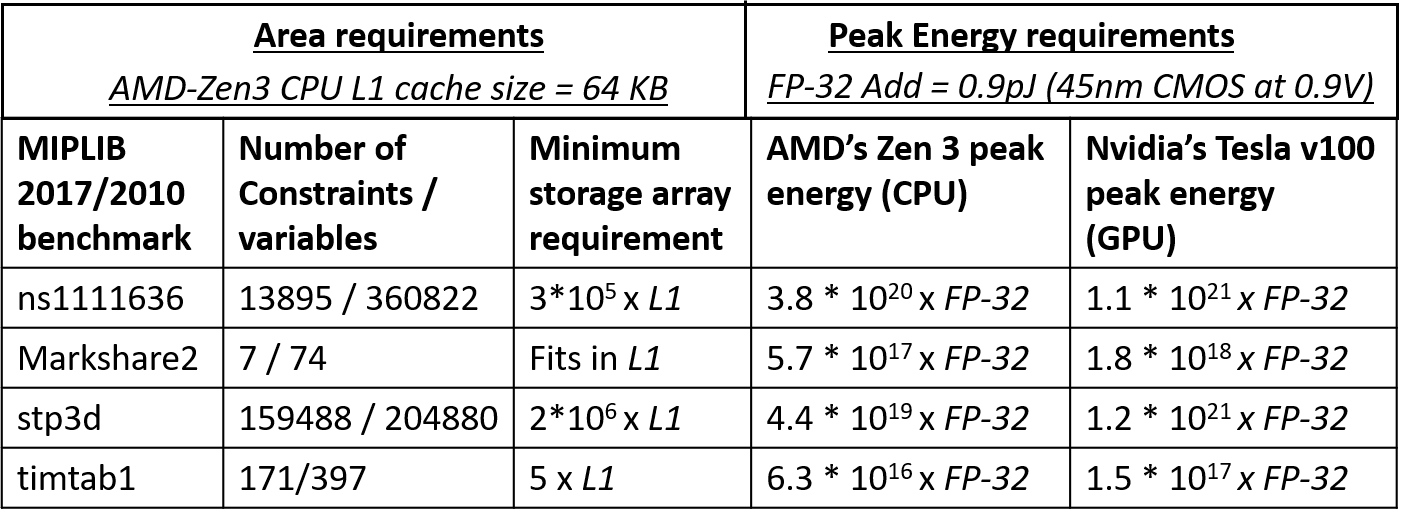}
\vspace{-2em}

\caption{
\textbf{\underline{ILP area}} -- The required memory array size is 10--200 GB in a few benchmarks, incurring frequent off-chip memory accesses.
\textbf{\underline{ILP energy}} consumption is in the order of $10^{6}$ Joules ($10^{17}$--$10^{20}$ times FP-32 add), because of data movement overhead and increased computational costs.
}

\label{Memory_energy}
\vspace{-1em}
\end{figure}

Linear Programming (LP)\cite{LP}\cite{LP_book} is an essential mathematical tool used to analyze a variety of optimization or feasibility problems, with the solution to these problems deduced based on a set of linear constraints. A variant of linear programming called integer linear programming (ILP), adds complexity by restricting the solution space to a set of integers. 
\begin{figure}[t]
\centering
\includegraphics[width=\linewidth]{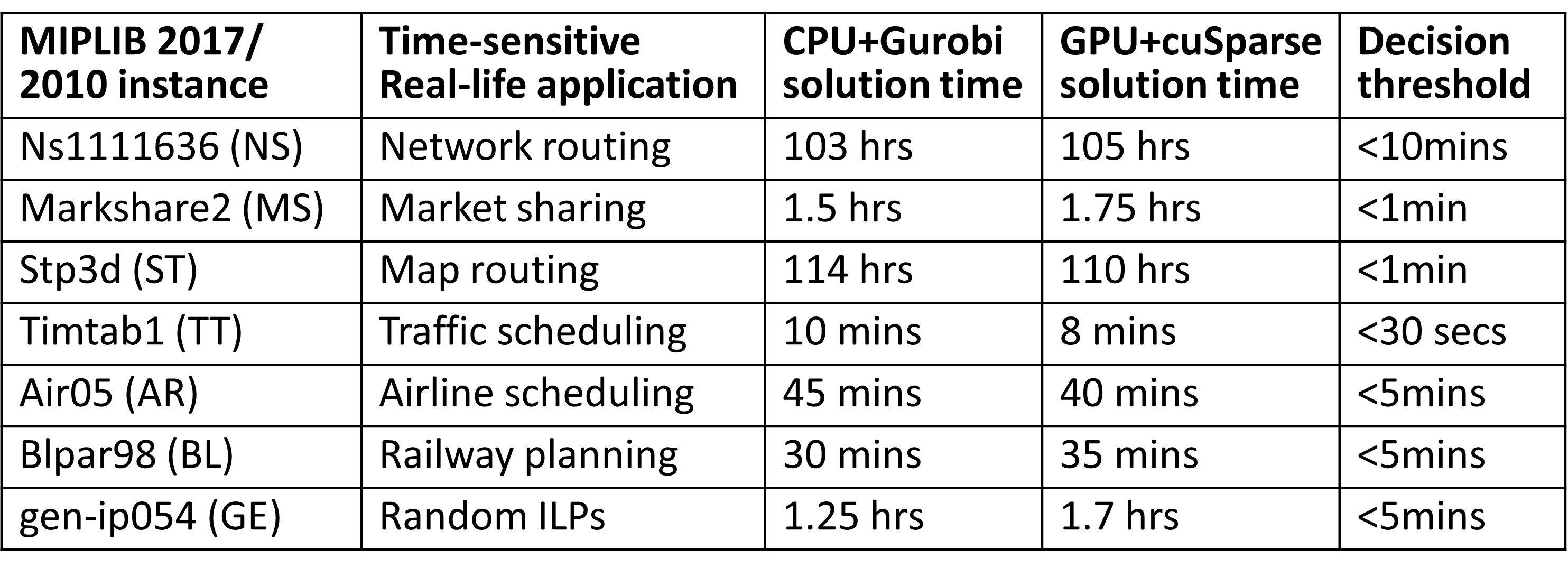}
\vspace{-2em}
\caption{\textbf{ILPs on CPUs and GPUs do not converge at the solution within the decision threshold time for time-sensitive real-life applications.}}
\label{Motivation_table}
\vspace{-1.5em}
\end{figure}
Solving ILP problems has been historically challenging~\cite{CPUs_ILP, Solvers_2020}.
\textbf Firstly, real-world ILP applications, like network routing, stock investments, traveling salesman, and emergency dispatch scheduling \cite{Solvers_2020}, demand timely decisions. Optimizers like Gurobi \cite{gurobi} and CPLEX \cite{CPLEX} use data patterns and multi-threading on CPUs for precise solutions. Koch et al. \cite{CPUs_ILP} show many ILP executions take tens of hours, even on 4000-8000 cores. While GPUs are an option, dataset sparsity (65-99\%) poses a challenge \cite{guorobi-blog}.
Execution times of state-of-the-art optimizations on CPUs and GPUs, as shown in Fig.~\ref{Motivation_table} for selected applications from MIPLIB 2017 \cite{MIPLIB_2017}, significantly surpass the decision threshold time for time-critical applications, even when leveraging libraries like cuSparse for GPU-based sparse problem execution.
\textbf{Secondly}, the problem size can be as large as 10\textsuperscript{5} variables and constraints, Fig.\ref{Memory_energy} indicates the storage array size required to store the constraints/variables as a function of the size of the L1 cache in Zen3 CPU. The minimum memory array storage required for storing the input constraints/variables (Fig.\ref{Memory_energy}), is in the range of 10-200 GB in a few instances. The modern-day on-chip memory size is in the range of KB-MB, implying that there is an increased storage requirement. Furthermore, most applications do not fit on-chip, requiring frequent off-chip memory accesses, costing performance/energy overhead. 

\textbf{Finally}, the energy to converge at the solution, 
can be extremely high ($\sim$ 10\textsuperscript{6} Joules), when MIPLIB benchmarks are executed on a CPU, because of data movement overhead and computational costs with large-sized sparse ILPs. This is 10\textsuperscript{17-20} times higher than a simple FP-32 addition (measured in 45nm technology \cite{Horowitz}) in CPUs and GPUS.

Domain-specific accelerators show promise for large problems with high runtimes and energy demands \cite{tpu-v4} \cite{tpu-v1} \cite{Cosa} \cite{SARA}, but solutions without dedicated accelerators are preferable. While the dot product in ILP benefits from parallelism, sparsity and control-intensive tasks present challenges. Additionally, moving large data between the host CPU and accelerator is a challenge.
Our solution involves a near-memory architecture integrated into the CPU, leveraging L1 cache for compute and minimizing data movement overheads. We allocate small area for dedicated sparsity-aware peripheral logic near L1 cache to handle control-intensive parts of ILP. The major features of SPARK are:

\begin{itemize}
   \setlength\itemsep{0em}
       \item \textbf{\underline{Sparsity awareness}:} A sparsity-aware algorithm is proposed to alleviate insignificant computes along with the ability to perform useful compute leveraging the high throughput of PIM, leading to energy efficient compute.

     \item \textbf{\underline{Reconfiguring CPU components}:} Spark is tightly integrated into the CPU pipeline, and reuses existing CPU components such as L1 cache, to accelerate ILP. 
    \item \textbf{\underline{Reuse-awareness}:} Identification of computational patterns in ILP algorithms to enable reuse of near-memory logic across different compute engines.
        \item \textbf{\underline{Near-cache architecture}:} The fine-grain near-memory dot product compute maximizes parallelism, while the coarse-grain dedicated hardware ensures high performance for control-intensive tasks, minimizing data movement  through in/near-memory computing.

    \item \textbf{\underline{ILP Speedup}}:  Spark achieves 15x/20x performance, and 152x/740x energy  improvements over AMD's Zen3 CPU/Nvidia's Tesla v100 GPU for sparse ILPs.

     \item \textbf{\underline{Broad applicability - LP Speedup}}: Spark is also suitable for LP problems in addition to ILP problems, and  results in 5-7x/150-180x performance/energy improvement over CPU/GPU in dense LP, and 7-17x/103-250x performance/energy improvement over CPU/GPU in sparse LP. 
    
\end{itemize}

\section{BACKGROUND}

\subsection{Integer linear programming (ILP) formulation}
Linear programming\cite{Data_Driven_library} (LP) solves optimization problems with non-negative solutions. Integer linear programming (ILP) \cite{ILP_solver, Acad_ILP} (special case of LP) restricts solutions to integers \cite{MIP}, leading to exponential time complexity.
 \par \textbf{ILP constraints} are represented by (i) a 2D matrix (C) with rows/columns equal to the number of constraints/variables, and (ii) a vector (D) with rows equal to the number of constraints. Additional constraints like X\textsubscript{j} $\geq$ 0 and X\textsubscript{j}$\in$ Z restrict the solution to non-negative integers \cite{Symmetric_ILP, Symmetric_ILP_1}.
 \par Thus, the general form of \textbf{optimization} version of ILP problem is $Optimize F(X)=\sum_{j=1}^{N} A_\mathrm{j}*X_\mathrm{j}$. The general form of the feasibility version of ILP problem is B = $\sum_{j=1}^{N} A\textsubscript{j}*X\textsubscript{j}$, where N is the number of variables, represented as X\textsubscript{1-N} in a system of linear equations. ILP's solutions satisfy these constraints: $i) C*X \leq D ; ii) X_\mathrm{j} \geq 0 ; iii) X_\mathrm{j} \in Z $
\begin{figure}[t]
\centering
\includegraphics[width=\linewidth]{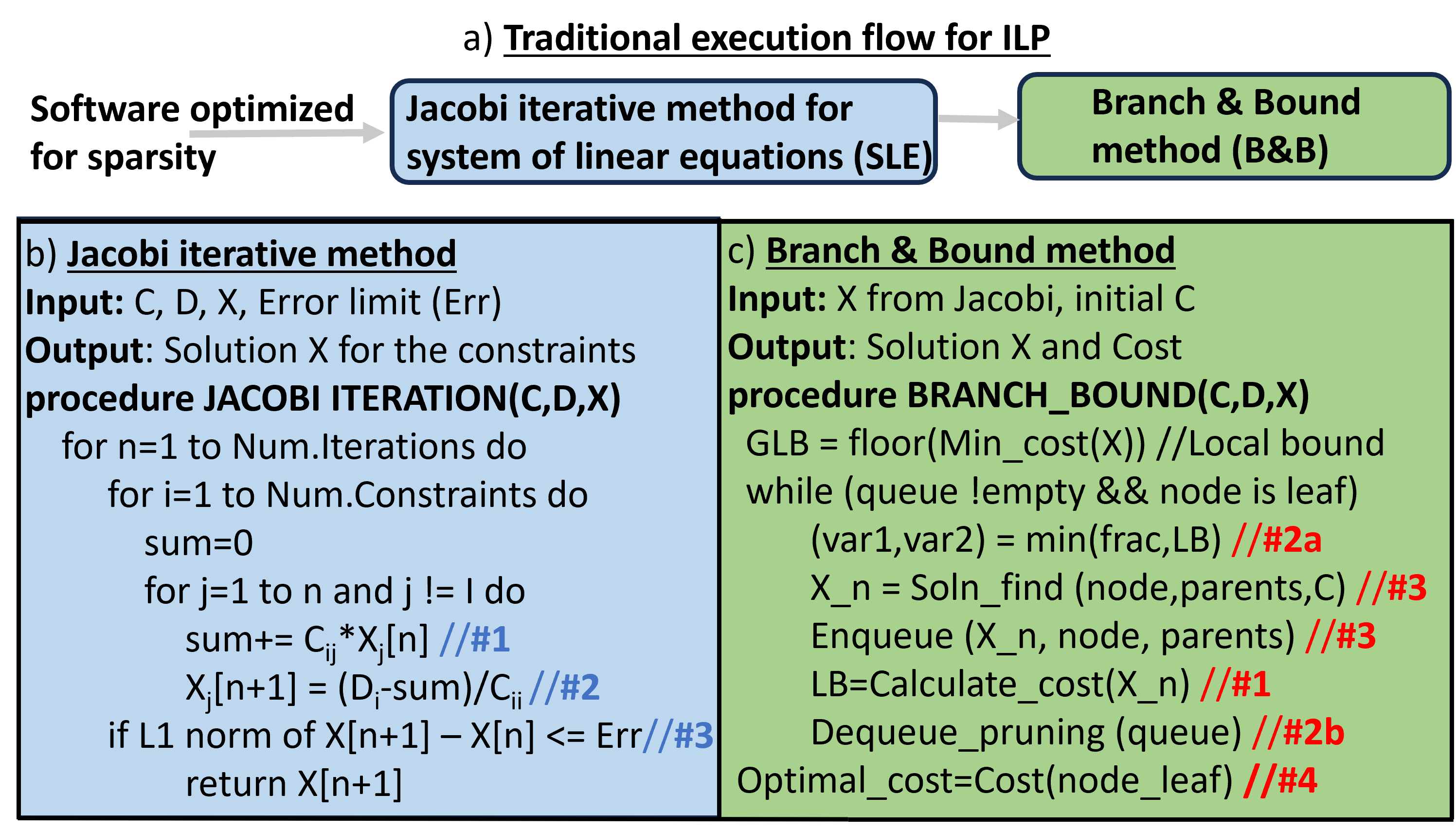}
\vspace{-2em}
    \caption{ \textbf{a) CPU/GPU-based ILP execution involves software optimization of sparsity, followed by SLE, B\&B methods. Pseudocodes for b) Jacobi iterative method (SLE) and c) Branch and Bound (B\&B).   }}
\label{SLE_BB_algo}
\vspace{-1em}
\end{figure}

\subsection{Traditional execution flow for solving ILPs}
Direct algorithms \cite{Simplex, Nelder-mead, Simplex_1} excel with smaller constraint sets, while iterative ones handle real-time optimization with complex constraints. Iterative methods like Jacobi and Gauss-Seidel \cite{Jacobi_iterative, Jacobi, Gauss-seidel} find optimal solutions, relaxing integer constraints, and Branch and Bound (B\&B) \cite{B&B_ILP, B_B_random} refines these to integer solutions. Fig.~\ref{SLE_BB_algo}.a shows ILP flow, with sparsity-optimized software executing SLE followed by B\&B. MIPLIB 2017 benchmarks reveal sparsity levels of 65\%-99\%.

\subsection{Jacobi method for system of linear equations (SLE)}
This iterative numerical technique solves a system of linear equations. It works with input constraints (matrix C), variables (X), and constants (D). i) \textbf{Initial Approximation}: Begin with a randomly initialized solution vector X. ii) \textbf{Iterative Updates}: Refine the solution iteratively by updating each variable. Each iteration computes new variable values based on the old values of all others. In Fig.~\ref{SLE_BB_algo}.b, \#1 and \#2 find updated X using MAC, subtraction, and division. iii) \textbf{Convergence criterion}: The process continues until a convergence criterion is met, which occurs when changes in X between iterations become sufficiently small (L1 norm). Label \#3 in Fig.~\ref{SLE_BB_algo}.b checks if L1 norm is less than a predefined error limit.

\begin{figure}[t]
\centering
\includegraphics[width=\linewidth]{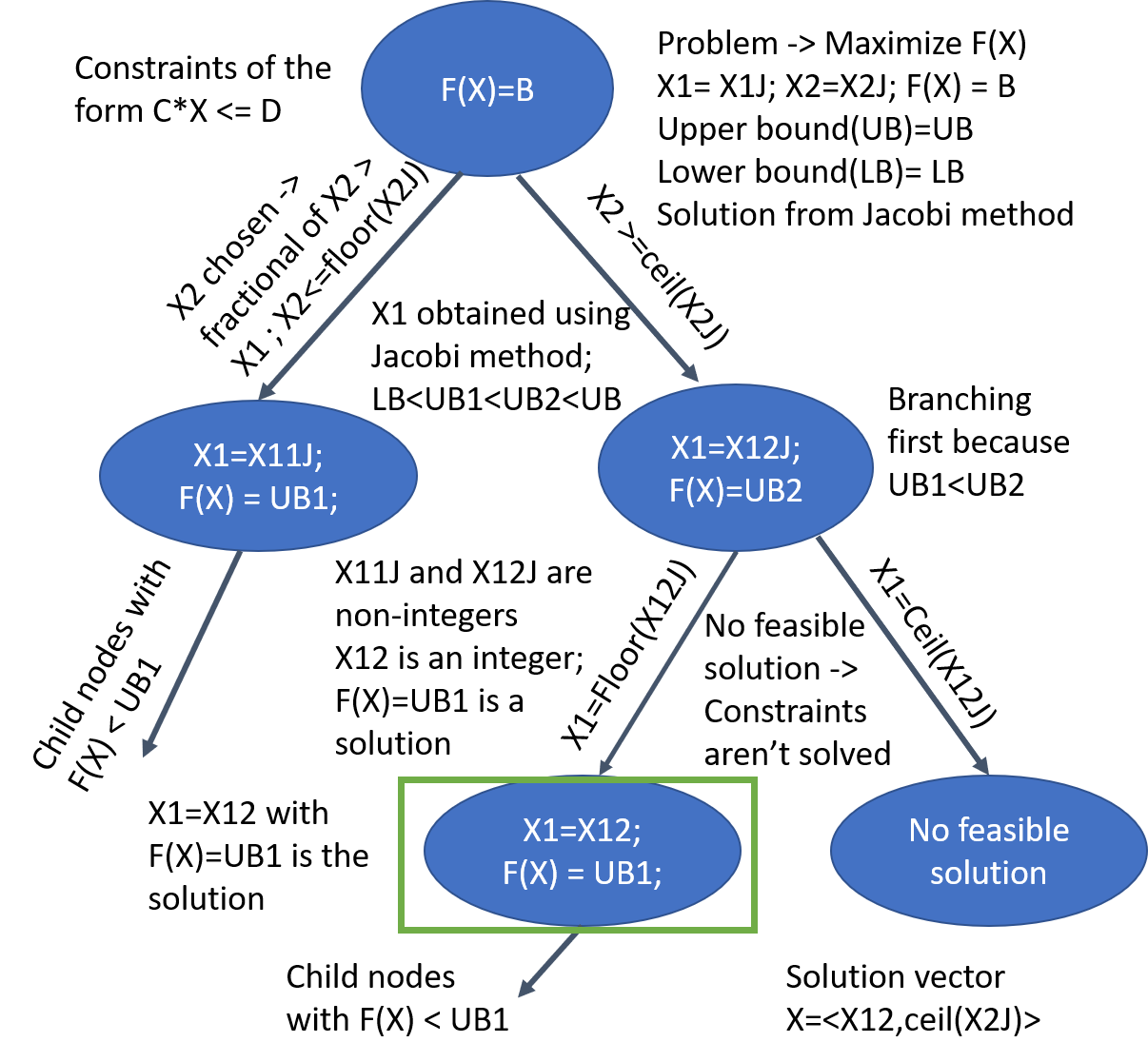}
\vspace{-2em}
\caption{\textbf{{Example of branch and bound algorithm for an optimization version(maximization problem) of ILP. The initial solutions from Jacobi iterative method are X1J, X2J for a system of linear equation with 2 variables.The obtained X vector solution after completing branch and bound process is $<$X12,ceil(X2J)$>$ }}}
\label{Branch_And_Bound}
\vspace{-1em}
\end{figure}
\subsection{Branch and Bound (B\&B)}
B\&B, after Jacobi, finds the optimal integer solution by generating child nodes for each parent node and selecting a branching variable based on the node's established bounds.
B\&B for minimization (min) problem involves steps shown in Fig.~\ref{SLE_BB_algo}.c: i) \textbf{Branching start}:  The lower bound (LB) is initialized to the ceil of $\sum_{j=1}^{n}A_\mathrm{j}*X_\mathrm{j}$. Global UB is the value of F(X) when X is the ceil of the obtained solution from SLE (\#1 in Fig.~\ref{SLE_BB_algo}.c);  ii) \textbf{Branching node/variable}: 
 Variable/node with the least fractional/local LB part branches first (\#2a in Fig.~\ref{SLE_BB_algo}.c); iii) \textbf{Branching continuation}:  C matrix updates each iteration to include new constraints from B\&B, and the updated X matrix is identified at each branch with the local LB of each node (\#3 in Fig.~\ref{SLE_BB_algo}.c). iv) \textbf{Branching complete}: Pruning occurs in four ways: (a) The solution vector X from SLE consists of non-negative integers. (b) When the local LB equals the global UB, and other local LBs are less than the global UB. (c) F(X) is identical across leaf nodes, with at least one node having non-negative integers. (d) F(X) is the same across some branches, with at least one valid solution and others infeasible (\#2b in Fig.~\ref{SLE_BB_algo}.c).
\subsection{Mathematical explanation of Branch and Bound}
The branch and bound(B\&B) algorithm is a popular technique that relies on breadth-first search(BFS) to converge at a solution for NP-hard problems like combinatorial optimization, ILP. However, the unique solution in the case of feasibility version of ILP limits their applicability to B\&B. The 3 major properties of this technique are: 1) The solution space (integers in case of ILP) form a rooted tree, with the root being the complete set of integers 2) The overall convex solution space can be partitioned into smaller subsets and breadth first search of smaller subsets leads to global optimal solution 3) Takes advantage of linear programming relaxation technique i.e. it eases out the constraint that the solution needs to be integer to identify the optimal integer solution. \par At the start of branching process, B\&B algorithm decides the branching variable and the node to branch from based on bounds, and then updates the bounds, decision on continuing at the end of the process. This results in a binary tree-like structure with each node defined by the upper/lower bounds. In this work, we choose the variable with the highest fractional part as the branching variable, however this choice can be made using other heuristics like choosing a random variable, choosing the variable with maximum value. For example, we assume that an initial set of solutions obtained from Jacobi iterative method as vector X=$<$X1,X2..Xn$>$ where n is the number of variables. The branching in Fig.\ref{Branch_And_Bound} is initiated from X2 node because the fractional part of X2 is greater than fractional part of X1. \par The process of bounding can be summarized as evaluating the possible solutions for equation \ref{eq:constraints} between the lower and upper bounds(LB/UB), with making modifications to the UB/LB at each branching node and keeping the LB/UB globally fixed in maximization/minimization problem. Incase of a maximization/minimization problem, the initial upper/lower bound is chosen as the floor/ceil of $\sum_{j=1}^{n}A_\mathrm{j}*X_\mathrm{j}$ respectively.  However, the lower/upper bound of maximization/minimzation problem is chosen as the value of F(X) when X vector consists of floor/ceil of the all the vector elements, respectively. Fig.\ref{Branch_And_Bound} shows the updated upper bound value for F(X) at every branching node and it must be noted that all the child nodes for a parent node have upper bound values lesser than F(X) making F(X) the upper bound across all the child nodes. The node with a higher/lower upper/lower bound is branched first in maximization/minimization problem. Fig.\ref{Branch_And_Bound} shows that X1=X12J is branched prior to X1=X11J because F(X)=UB1 is lesser than F(X)=UB2. Furthermore, the value of X1=X11J or X1=X12J for these nodes is identified by re-computing the initial constraints matrix with the additional constraints being X2$\leq$floor(X2J) or X12$\geq$ceil(X2J) respectively. The process of stopping the branching, also known as pruning can be accomplished in 4 means: 1) Branching is not initiated if the obtained solution vector X belongs to a set of non-negative integers. 2) If one of the branching nodes hits the global lower/upper bound in maximization/minimization problem implying that these bounds are the solutions for the optimization problem. 3) The value of F(X) is identical across all the branches with X vector of atleast one node contains non-negative integers. 4) F(X) is same across few branches with atleast one integer solution and the other branches have no feasible integer solutions. Fig.\ref{Branch_And_Bound} shows that the F(X)=UB1 is the maximum upper bound obtained in both the branches- X2$\leq$floor(X2J)/ X12$\geq$ ceil(X2J), with one of the branches having an integer solution. Thus, no child nodes are created for the node, whose branches are pruned and the leaf nodes that have the potential to initiate branching are called active nodes. No feasible solution implies that there can be no child nodes to that particular node in the tree. The solution vector for ILP in   Fig.\ref{Branch_And_Bound} is X1=X12,X2=ceil(X2J). \begin{align}\label{eq:constraints}
   Constraints <- 
\begin{cases}
    C*X \leq D   \\
    X_\mathrm{j} \geq 0 \\
    X_\mathrm{j} \in Z 
\end{cases}
\end{align} 


\subsection{Memory operations}
Memory is organized into multiple subarrays, with each subarray containing rows and columns of bitcells storing data, a row of sense amplifiers (row buffer), for sensing data stored in a row of the subarray. This memory organization can be seen in caches, made of 6T/8T Static Random Access Memories(SRAM), 1T1C/3T1C embedded Dynamic Random Access Memory(eDRAM) bitcells. Each bitcell in the 6T SRAM/1T1C subarray consisted of a shared read/write port. Write is performed by turning on the word line(WL) of a row and writing data using Bit Line(BL) of a column. The read of bitcell contents is performed by precharging the BL, turning on the WL of a row and sensing BL discharge using row buffer. The disadvantage with shared read/write ports is the inability to perform back-to-back write and read accesses (commonly seen in caches) because 'write' is followed by 'precharge' before performing a 'read'. For this, usage of memory with decoupled write and read ports(8T SRAM, 3T1C eDRAM)\cite{8T_SRAM}\cite{8T_SRAM_1}\cite{eDRAM_1}, with write performed using W(Write)WL and WBL, read performed using R(Read)WL and RBL was proposed. The major advantage is that precharge is done in parallel with write, ensuring that write after read is a 2 cycle operation. 

\section{MOTIVATION}
\subsection{Understanding 3C criterion for ILP execution}
To explain SPARK's necessity, we highlight why existing accelerators are unsuitable for ILP execution. ILP involves more than just matrix operations, requiring the \textbf{3C criterion}: \textbf{C1} - Dot-product intensive operations in SLE, \textbf{C2} - Handling sparse constraints, and \textbf{C3} - Managing control flow tasks like L1 norm in SLE, B\&B. Existing accelerators typically fulfill at most two of these, not all.


\begin{figure}[t]
\centering
\includegraphics[width=\linewidth]{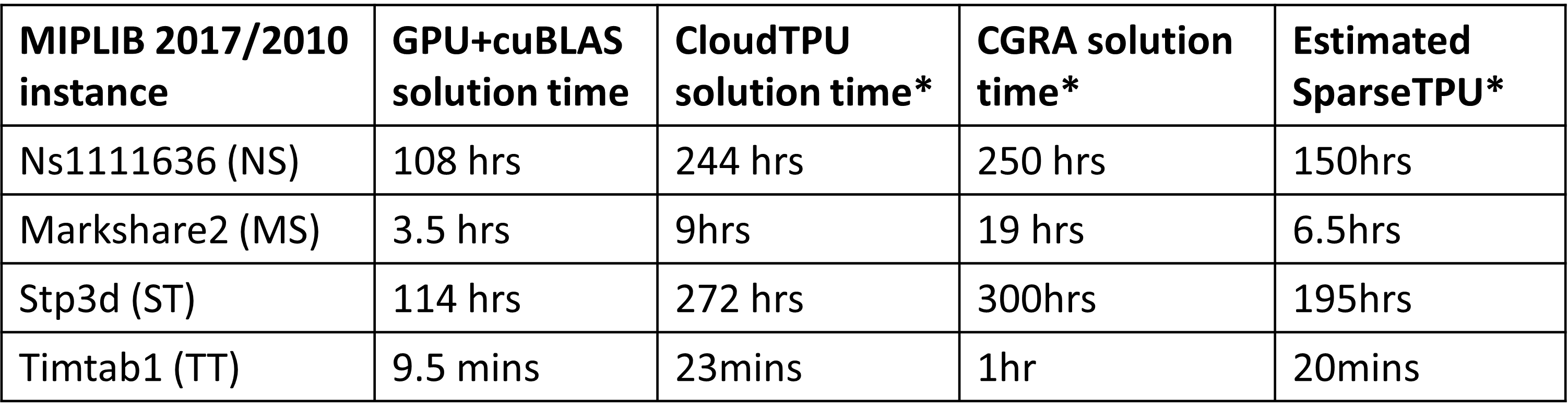}
\vspace{-2em}
\caption{{\textbf{Experiments with TPUs/CGRAs show unacceptable solution times (in hours), even at reduced accuracies (* indicates 98\% of CPU accuracy is achieved).}}}
\label{Comparison_new_accelerators}
\vspace{-1em}
\end{figure}
\subsection{Shortcomings of Linear Algebra/Tensor accelerators}
Traditional linear algebra accelerators handle dot-product intensive operations such as matrix-matrix/matrix-vector multiplications in convolution, fast-fourier transform for signal processing. These are mainly classified as application-specific integrated circuits (ASIC) like TPU, Bison-e, and reconfigurable architectures like CGRA, Transmuter \cite{Transmuter}, etc.
\par Experiments with TPUs for ILP workloads show they underperform compared to CPUs/GPUs, which already struggle to meet decision thresholds (Fig.\ref{Motivation_table}). Results in Fig.\ref{Comparison_new_accelerators} align with CloudTPU documentation \cite{CloudTPU}, highlighting TPUs' inefficiency for branching and sparse operations. Additionally, solution accuracy on TPUs is only 98\% of that on CPUs, compromising both accuracy and solution time.
\par \underline{Bison-e} \cite{Bison-e}, an ASIC optimized for generic integer linear algebra applications, uses binary segmentation for matrix-matrix/vector multiplication, addressing characteristic C1. However, Bison-e currently lacks control for handling sparse, control-flow intensive operations, unsuitable for ILP, and framework/compiler support is still in nascent phase.
\par Coarse-grained reconfigurable architectures (CGRA) are valued for their reconfigurability, mapping problems to data-dependency graphs (DDG). However, they rely heavily on compilers and perform poorly on ILPs with conditional statements \cite{CCF}. Data sharing between the CPU and CGRA is complex with current frameworks, limiting their suitability for real-world ILPs (Fig.\ref{Comparison_new_accelerators}). We also observed sparsity and control flow issues causing underutilization of processing engines.

\par \underline{Transmuter} \cite{Transmuter} uses a reconfigurable data-flow model (like CGRA), adaptable memory, and cross-bar arrays, allowing kernel computation with varying arithmetic intensity. However, it has drawbacks: i) Only 2x speedup over CGRA, insufficient compared to CPUs/GPUs. ii) Frequent data movement from the host CPU. iii) High integration overhead for control engines resembling local CPUs. iv) Unclear energy cost for clocking local control units. v) High reconfigurability overhead for sparse/branching workloads.
 \begin{figure}[t]
\centering
\includegraphics[width=\linewidth]{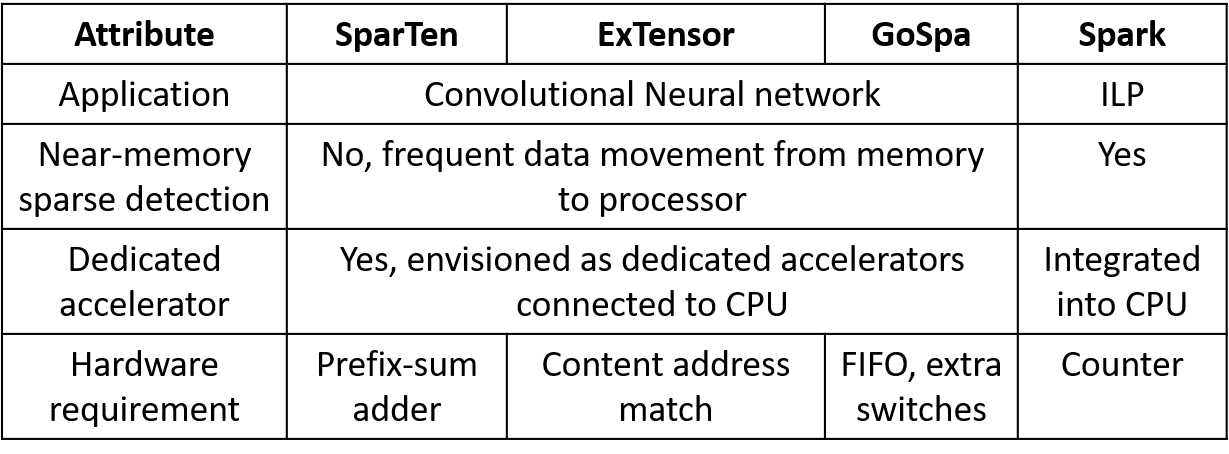}
\vspace{-2em}
\caption{\textbf{\underline{Sparsity aware}: In contrast to sparsity-aware accelerators for CNNs like SparTen \cite{SparTen}, ExTensor \cite{Extensor}, and GoSpa \cite{GoSpa}, Spark is a near-memory sparsity-aware accelerator using counters.}}
\label{Sparsity_awareness}
\vspace{-1em}
\end{figure}
\subsection{Shortcomings of existing sparsity-aware accelerators}
\par Numerous sparsity-aware accelerators, dedicated to machine learning, are unsuitable for solving ILPs due to their i) inability to handle control-flow-intensive operations. ii) Offloading such operations to the host requires periodic host-accelerator data movement, impacting performance/energy, as shown in Fig.\ref{Sparsity_awareness}. 
\par \underline{Sparse-TPU} \cite{Sparse_TPU} addresses sparsity by leveraging static sparsity in weights, allowing offline encoding of data and transformations for arranging them as structured dense computations. i) Sparsity-aware TPUs incur a fundamental overhead. Data must be pre-processed to ensure that non-zero elements access the correct index/PE block across processing engines aligned in a mesh style, adding extra overhead. \cite{Survey} ii) The proposed 2D tensor matrix approach requires adaptation to modern TPUs at a 3D level. Column packaging may not achieve optimal density for a 3D matrix. iii) Software-based sparsity encoding/detection is slower compared to extremely parallel hardware-based sparsity detection. iv) Sparse TPUs, at best, exhibit characteristics of C1 and C2, but struggle with control-flow-intensive operations, leading to frequent data movement cost and rendering them unsuitable for ILP operations. v) A reported 16x speedup aids SLE but not L1 norm or B\&B, falling short of CPU/GPU performance levels. Impact on solution accuracy remains uncertain.
\par \underline{EIE} \cite{EIE} is a specialized DNN hardware accelerator that employs Deep Compression for network pruning and utilizes a dedicated pipeline for matrix-vector multiplication. Other sparsity-aware accelerators, such as \underline{SparTen} \cite{SparTen} and \underline{ExTensor} \cite{Extensor}, require components like prefix-sum adders and content address match, designed specifically for CNN. These cannot handle B\&B or L1-norm operations due to their control logic limitations, requiring offloading to the host CPU, increasing data movement and energy costs. To modify them into ILP accelerators, additional structures like queues, subtractors, and dividers would be needed for B\&B and L1-norm, effectively requiring a separate accelerator. This is because they are designed only for sparse dot-product computations, and their existing structures can't be reused for B\&B/L1-norm.

\subsection{Shortcomings of Ising/Boltzmann accelerators}
\underline{Boltzmann/Ising accelerators} map NP-complete Combinatorial Optimization Problems onto an Ising graph, where spins (S) and interaction coefficients (IC) represent variables and constants, respectively. Existing Ising accelerators face limitations for ILPs: i) Binary-valued spins limit their applicability to binary ILPs, unlike real-life applications. ii) Most Ising accelerators like \cite{Shanshan_Ising}\cite{CIM-Spin}\cite{SACHI} rely on  Hamiltonian energy being represented as a quadratic formulation of product between S and IC, suitable for PIM compute. However, binary ILPs warrant a different Hamiltonian energy (mentioned in \cite{Ising}), not captured by the existing Ising accelerators. 

\begin{figure}[t]
\centering
\includegraphics[width=\linewidth]{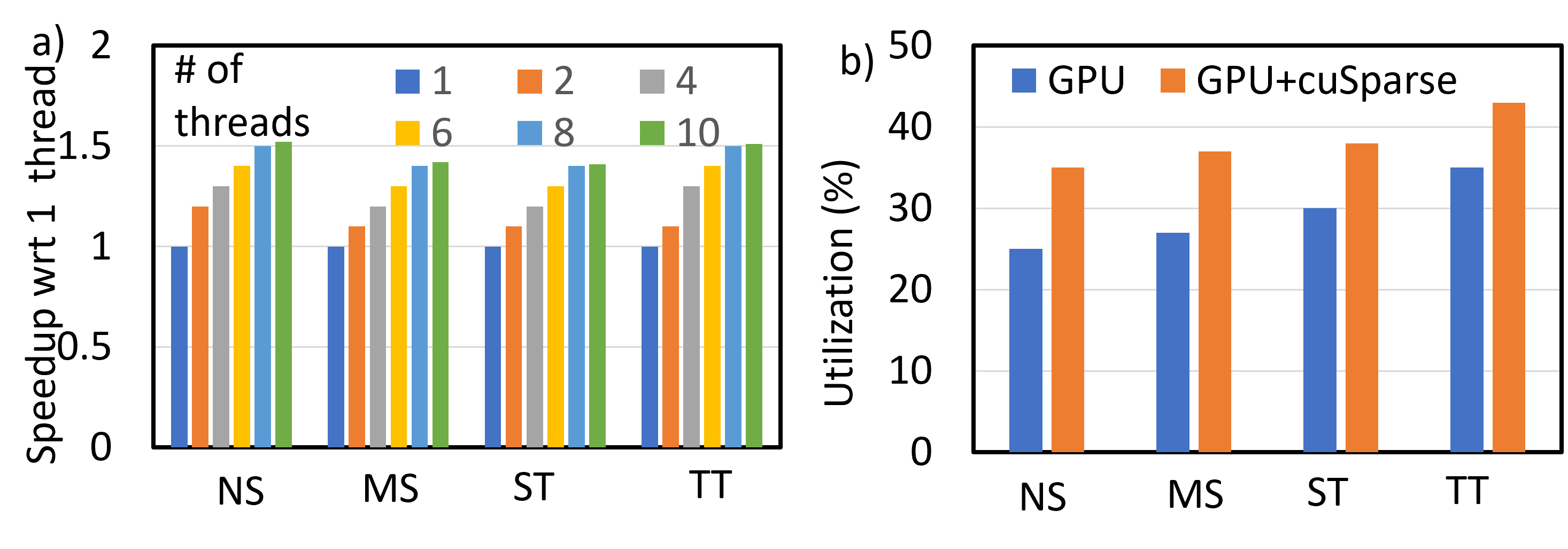}
\vspace{-2em}
\caption{\textbf{a) ILP on CPU - Performance saturation with increasing number of threads suggests hardware bottlenecks like limited throughput and high data movement. b) \textbf{ILP on GPU} - GPU utilization with/without cuSparse is less due to sparsity and thread divergence}}
\label{CPU_problems}
\vspace{-1em}
\end{figure}
   
\begin{figure}[t]
\centering
\includegraphics[width=\linewidth]{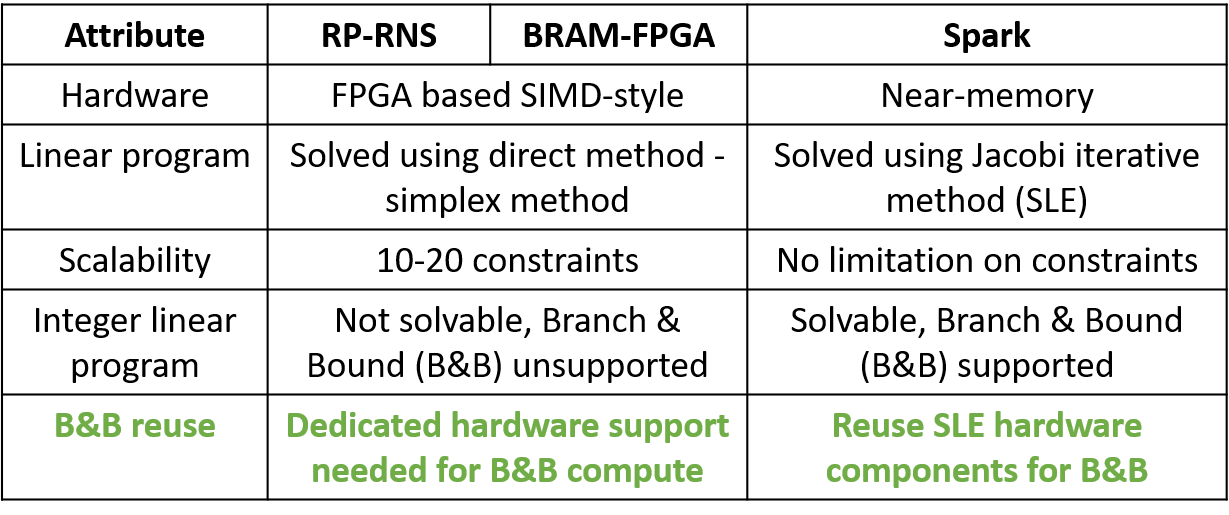}
\vspace{-2em}
\caption{\textbf{\underline{Reuse aware}: RP-RNS \cite{SLE_SOC} and BRAM-FPGA are \cite{FPGA_ILP} small-scale LP solvers, not capable of solving ILP, as Branch and Bound (B\&B) is unsupported. Spark solves LPs by using Jacobi iterative method (SLE) with no constraints limitation and solves ILPs by reusing components.} }
\label{Reuse_aware}
\vspace{-1em}
\end{figure}
\subsection{Shortcomings of CPU/GPU based execution}
While {\underline {multicore CPUs} can be used for ILP solving, Koch et al. ~\cite{CPUs_ILP} presented the inadequacy of solving ILPs on multicores, 
Fig.~\ref{CPU_problems}.a suggests that increasing threads do not scale performance
(\cite{CPUs_ILP}). 
CPUs, using sparsity-optimized software like Gurobi rely on Von Neumann compute. \par {\underline {GPUs} can solve ILPs, but the sparsity poses challenges~\cite{guorobi-blog}}. 
Data is transferred from the CPU's Data (D) cache via shared memory to GPU cores, incurring data movement overhead leading to energy bottleneck. Fig.~\ref{CPU_problems}.b shows the under-utilization in GPUs with/without cuSparse, because of sparsity and thread divergence \cite{gurobi}\cite{Thread_divergence_GPU}, negatively affecting efficiency \cite{Branch_div} \cite{Centaur}. GPU+cuSparse/cuBLAS (Fig.\ref{CPU_problems}) were compared, with the former outperforming the latter. Subsequently, results from cuSparse alone are presented.

\subsection{Challenges in existing B\&B accelerators}
\par Software optimizations to accelerate B\&B have been proposed~\cite{Thread_divergence_GPU, Fast_branch_bound}, while no hardware accelerators for B\&B exist. To reduce thread divergence in GPU-based B\&B, Chakroun et al. \cite{Thread_divergence_GPU} used an entirely software-driven optimization for executing branches in parallel. 
The authors of \cite{Fast_branch_bound} propose a fast algorithm for optimal sub-problem identification in feature selection. However, these suffer from sparsity, necessitating periodic host-GPU interaction and unnecessarily lengthening the time of short threads, as shown in Fig.\ref{Reuse_aware}.  

\begin{figure}[t]
\centering
\includegraphics[width=\linewidth]{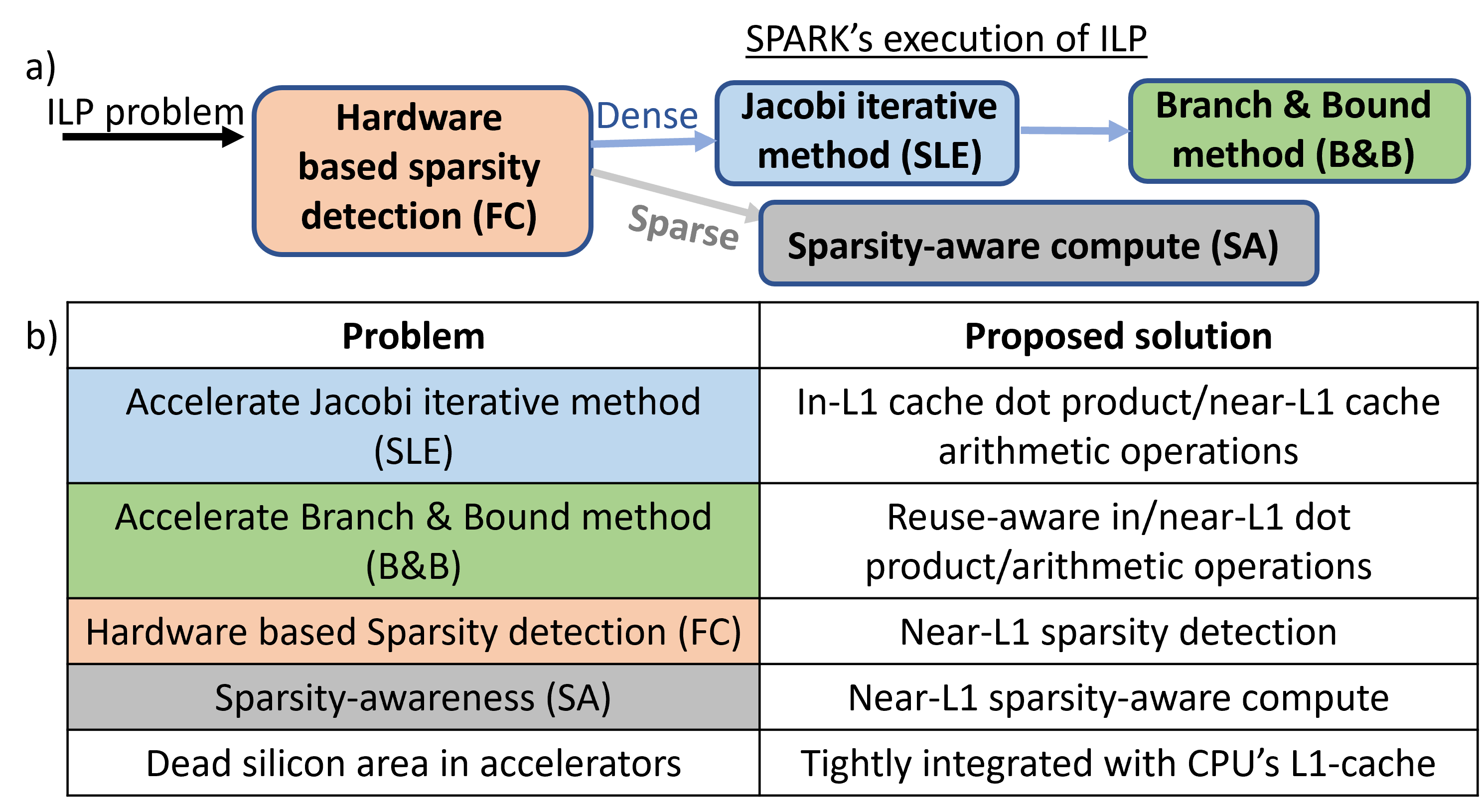}
\vspace{-2em}
\caption{\textbf{a) SPARK uses hardware-based sparsity detection, execution based on sparsity. b) Problems in solving ILPs and proposed solutions}}
\label{Architecture_ILP}
\vspace{-1em}
\end{figure}

\subsection{Challenges in existing Linear Programming accelerators} Prior research~\cite{SLE_SOC}\cite{RNS} investigates the use of a linear solver System on Chip (SoC) architecture that utilizes a Residue Number System (RNS) combined with residual processors (RP) functioning as SIMD (Single Instruction, Multiple Data) units. While this approach offers parallel computation benefits, RNS suffers from significant limitations, primarily its low dynamic range, which restricts its ability to handle problems with a wide range of values. This limitation hinders convergence, reduces performance, and makes it difficult to apply the system to problems with more than 10-20 constraints or larger-scale Linear Programming (LP) problems. Another study~\cite{FPGA_ILP} explores the use of FPGA for implementing the Simplex algorithm, leveraging block-RAM for data storage to speed up the optimization process. However, this FPGA-based solution faces scalability issues, particularly in terms of its ALUs, and it lacks the necessary support for solving Integer Linear Programming (ILP) problems using Branch-and-Bound (B\&B) methods.

 \subsection{Challenges in existing processing-in-memory accelerators} PIM  accelerators have been proposed for a wide range of computational problems~\cite{neuralcache} \cite{comefa} \cite{wave-pim} \cite{Compute_cache}, but they are found to be inadequate for the specific needs of ILP acceleration. One key issue is the use of a 1-bit adder on the column-lines, which proves insufficient for effectively addressing two critical aspects of ILP computation. First, it struggles to accelerate the sparse, control-flow-intense parts of ILP problems, where efficiency is key. Second, it does not provide sufficient throughput, preventing the near-memory logic from fully exploiting its potential. Finally, the existing PIM accelerators do not handle sparsity extremely well. For effective ILP acceleration, it is essential to handle control-flow intensive throughput-heavy sparse operations efficiently.

\section{THE SPARK ARCHITECTURE}
\subsection{Learning from shortcomings of prior approaches}
Traditional accelerators for linear algebra, sparsity, Ising/Boltzmann models, CGRAs, and ILP-related works struggle with sparsity, control-heavy operations, and data movement between host and accelerator. CPU and GPU optimizations outperform them, with CPUs leading. SPARK reconfigures L1 cache for dot-product compute, adding minimal near-L1 logic across CPU cores for hardware-based sparsity detection, SLE/B\&B execution (Fig.~\ref{Architecture_ILP}.a).


\subsection{SPARK's acceleration strategy satisfying 3C criterion}
SPARK's acceleration strategy (Fig.~\ref{Architecture_ILP}.b) satisfies the \textbf{3C criterion}:
(i) \textbf{C1}: Optimizes SLE compute throughput via in/near-memory arithmetic operations.
(ii) \textbf{C2}: Handles sparsity with algorithmic transformation and sparsity-aware compute.
(iii) \textbf{C3}: Accelerates B\&B by optimizing control flow and reusing SLE components.
(iv) SPARK is compact, energy-efficient, and easily integrates into SoCs.

\subsection{Choice of tightly integrated over dedicated accelerator}
 Integrating dedicated accelerators into modern SoCs is challenging due to i) accelerators increasing area overhead/cost, ii) handling control-intensive tasks suited for CPUs, adding overhead to throughput-focused accelerators, iii) high data movement cost between CPU and accelerator, even for small MIPLIB benchmarks, increasing power overhead, and iv) requiring a distinct programming model, unlike adding instructions to an existing ISA, which is adaptable for programmers.

 \begin{figure}[t]
\centering
\includegraphics[width=\linewidth]{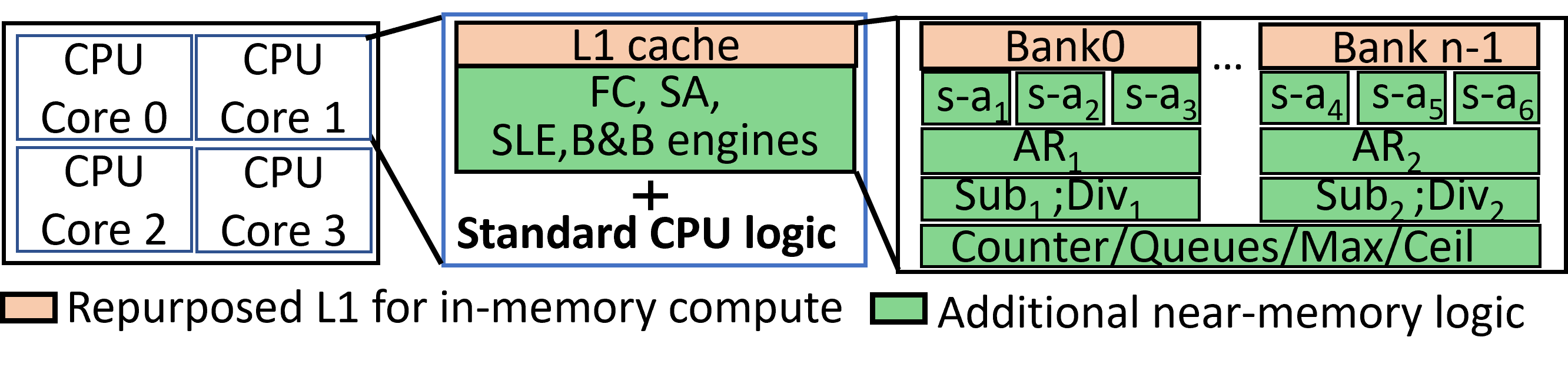}
\vspace{-2em}
\caption{ \textbf{Spark is realized by re-configuring L1 cache (orange) in CPUs for PIM along with minimal near-memory logic (green) shared among FC, SA, SLE, and B\&B engines. In an L1 cache with n banks, engines are realized using: shift-add (s-a\textsubscript{1-3}) at a finer granularity of 1 per 16 columns in a bank for 16-bit compute, adder reduction (AR\textsubscript{1}) for s-a outputs, subtraction (Sub\textsubscript{1}), and division (Div\textsubscript{1}) at a coarser granularity of 1 per bank. The counters/queues/max/ceil are shared across the cache.}}
\label{CPU_GPU_proposed}
\vspace{-1em}
\end{figure}

\subsection{SPARK's tightly integrated architecture}
SPARK reconfigures the L1 cache and the added near memory logic to realize the following engines: i) FC (Fetch/Control) engine for sparsity detection; 
ii) SA (Sparsity-aware) engine for sparsity-aware compute.
iii) SLE engine for SLE acceleration, 
iv) B\&B engine for B\&B acceleration. \par \underline{Micro-architecture overview}: Fig.~\ref{CPU_GPU_proposed} shows SPARK's architecture on a 4-core machine, with n banks of L1 cache per core and near-memory logic shared to realize different engines. The shift-and-add (s-a\textsubscript{1-3}) operates on partial dot products in memory, with 1 per 16 columns per bank for 16-bit compute, aligning with MIPLIB value ranges.
The adder reduction (AR\textsubscript{1}) of s-a outputs/subtractors(Sub\textsubscript{1})/dividers(Div\textsubscript{1}) are present at a coarse granularity of 1 per bank, with additional counters/queues shared across L1-cache. 
\subsection{SPARK's choice of units} SPARK's design focuses on selecting its computational units to minimize area while ensuring that throughput is maintained. In SLE compute, the primary arithmetic operation is MAC, crucial for updating variables by performing MAC across X and C vectors with all other variables (\#1 in Fig.~\ref{SLE_BB_algo}.a).  Because of the importance of the MAC operation in updating these variables, the SA units are designed to operate at a finer granularity to efficiently handle these operations in parallel. In contrast, subtraction and division operations are only needed after all the MAC operations have been computed. These operations require much less complexity and thus can be handled with a single 2-input operation, as illustrated in \#2 of Fig.~\ref{SLE_BB_algo}b). Therefore, these are present at a coarser granularity. Similar reasoning can be extended to proposed reuse-aware B\&B, sparsity-aware ILP compute.

  \begin{figure}[t]
\centering
\includegraphics[width=\linewidth]{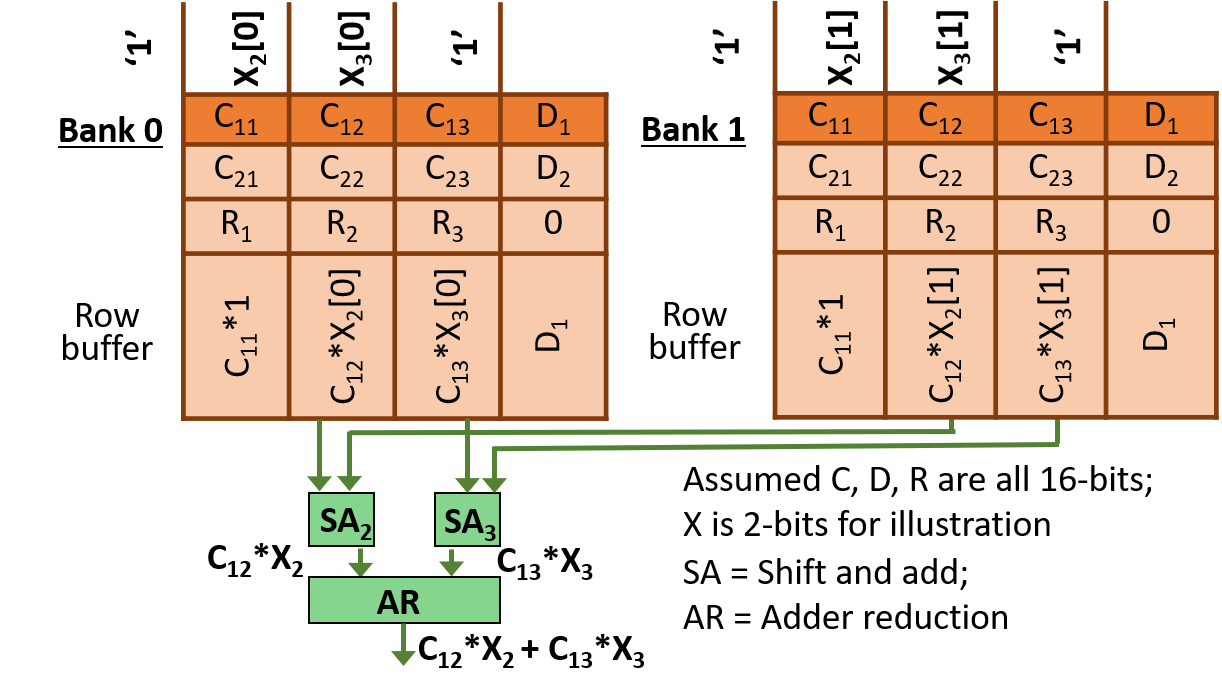}
\vspace{-2em}
\caption{ \textbf{In-L1 dot product compute followed by near-L1 accumulation, adder reduction for SLE solving while re-using row buffer/sense amplifier}} 
\label{SLE_engine_PIM}
\vspace{-1em}
\end{figure}
\subsection{SPARK - L1 cache reconfigured for compute}
\underline {Architecture}: The L1 cache in modern processors is typically organized into multiple banks and utilizes 8T SRAM bitcells (\cite{8T_SRAM}\cite{8T_SRAM_1} \cite{eDRAM_1}\cite{IGZO_CIM} \cite{8T_L1}\cite{8T_L1_2}\cite{GCN}\cite{TACO}\cite{NEM_GNN_arxiv}), which feature decoupled read/write ports. This configuration allows for efficient data access, as read and write operations can occur simultaneously without interfering with each other. SPARK stores constraint coefficients (C\textsubscript{ij}, D\textsubscript{i}) and cost function (R\textsubscript{i}) in L1 cache, reconfiguring it for compute using decoupled read port.

\par \underline{Example}:Fig.~\ref{SLE_engine_PIM} shows in/near-memory compute for SLE step \#1 for X\textsubscript{0} using PIM(L1)+SA+AR. Assuming C, D, R are 16 bits and X is 2 bits, C/D is replicated across 2 banks. The 1\textsuperscript{st}/2\textsuperscript{nd} banks handle the dot-product between C (in memory) and the 0\textsuperscript{th}/1\textsuperscript{st} bit (mapped to RBL) of X. Row-buffer stores the dot-product result. SA shifts and adds partial products, and AR reduces them to calculate C\textsubscript{12}*X\textsubscript{2}+C\textsubscript{13}*X\textsubscript{3}.

 \subsection{Circuit details for in-memory compute} L1 cache uses 8T SRAM with decoupled read/write ports (RWL, RBL, WWL, WBL) allowing read-after-write in 2 cycles, compared to 3 cycles for 6T SRAM \cite{6T_SRAM}, offering a performance benefit. These have decoupled read (RBL- read bit-line/RWL- read word-line) and write (WBL- write bit line/WWL - write word line) ports, wherein decoupled read-port is "reconfigured" for compute. All columns are computed in 8T SRAM array in parallel. \par Dot product compute is performed by storing coefficients onto the bit-cell (SN) and pre-charging RBL based on the value of the incoming variable (X) (Fig.~\ref{PIM}.a,b). For X='1'/'0', RBL is precharged to V\textsubscript{cc}/(V\textsubscript{cc}/2) respectively. For performing dot product between '1' (SN) and '1' (RBL), RBL discharges below V\textsubscript{cc}/2, while RBL is greater than V\textsubscript{cc}/2 in other cases. RBL discharges via read-port transistors marked in orange, and is sensed by repurposing the sense-amplifier logic to obtain dot product, without modifying array.


\subsection{Choice of L1 cache over last level cache}
This choice considered
factors like size, performance,
throughput, and energy efficiency for MIPLIB benchmarks. 
\par \underline{Observation}: $\sim$65\% of MIPLIB
2017 benchmarks, fit within a 128KB L1 cache. For 65\% of
workloads, L1 cache demonstrated: i) superior performance
with low access latency, ii) sufficient throughput accommodating the entire workload, and iii) lower energy demands due
to reduced data movement compared to constant CPU core
requests to LLC. 35\% of benchmarks exceeding
L1 capacity led to a trade-off analysis between L2 and L1 caches for energy efficiency (throughput divided by energy) in an 8MB LLC versus a 128KB L1 cache. \par \underline{Results}: Simulations show that masking data movement from L2-L1 outweighs LLC’s throughput benefits, resulting in 20-25x energy efficiency enhancement in MIPLIB benchmarks. L1 cache was selected, incorporating prefetching to reduce data movement latency while maximizing efficiency.
\begin{figure}[t]
\centering
\includegraphics[width=\linewidth]{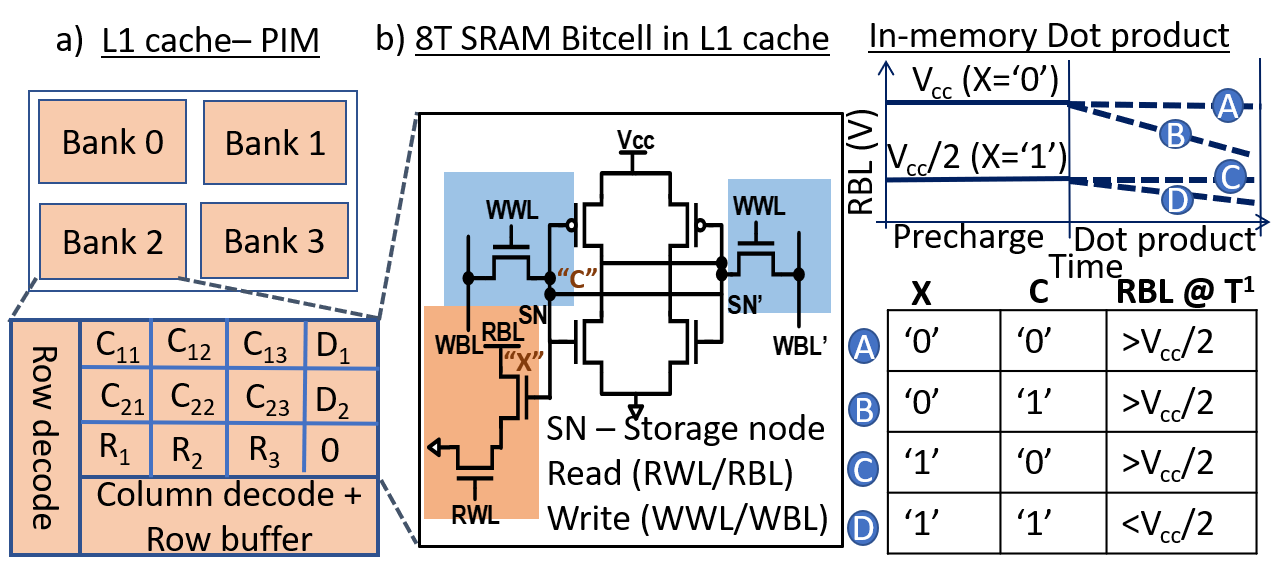}
\vspace{-2em}
\caption{\textbf{a) L1 cache, organized as banks, stores C and cost function (R) consisting of b) 8T SRAM bit cells with decoupled read (orange) and write ports (blue). A data-dependent precharge maps X onto RBL with C stored in bitcell. \textbf{Dot-product compute} between X and  C is identified by the value of RBL. RBL at T\textsuperscript{1} $>$ Vcc/2 =$>$ '0', RBL $<$ Vcc/2 =$>$ '1'.}}
\label{PIM}
\vspace{-1em}
\end{figure} 
\subsection{Prefetching for large workloads}
\underline{Idea}: For large workloads that exceed the capacity of the L1 cache, we have implemented a robust prefetching mechanism to effectively address performance and throughput bottlenecks that arise due to cache misses. This strategy takes full advantage of the highly structured and predictable nature of L1 cache accesses. We ensure the accesses to L1 cache start from top to bottom of PIM array, ensuring determinstic compute.
\par \underline{Timeliness/prefetch location}: Initiating sequential prefetching requests early enough ensures fill latency amortization in case of overflow. The choice of a sequential access order is particularly advantageous because the convergence iterations within the computation process are independent of the order in which variables are updated. This independence effectively eliminates the possibility of performance bottlenecks that might otherwise occur if variable update order were a limiting factor. Data is efficiently filled into the L1 cache via the CPU core's dedicated fill pipeline. The data filling process is optimized by replacing the least recently "computed" input constraint, which ensures that the cache is continually populated with the most relevant data for ongoing computations. This approach takes advantage of the L1 cache's ability to support simultaneous read and write operations to different indices, a feature made possible by the decoupled read and write ports in the cache architecture. 

 \subsection{Impact on traditional CPU workloads}  
We compared timing metrics with/without near-memory in load-store unit in a 2mm*2mm floorplan and 2ns clock latency in 45nm technology, post place,route, regarding i) increased gate depth affecting critical paths, ii) placement disruptions in conventional logic due to the near-memory logic, iii) latency, access ports, energy, capacity, associativity.\par{\underline{Gate depth}:} The timing of critical paths in the system remains unaffected by the proposed changes. Specifically, there are three key factors contributing to this: (i) Near-memory logic operates in a separate inactive pipeline during normal operations, and do not add gate depth. The typical critical path for reads involves setting up the read-index (virtual address in VIPT cache), wherein the address can be forwarded from execution units. This path remains unaffected by SPARK. (ii) In the case of processing-in-memory (PIM), no significant modifications are made to the array itself. The only change is the introduction of a 2:1 multiplexer on the periphery, which is used to perform dot-product computations. This multiplexer delay is absorbed with no latency impact (iii) Fill datapath, which gets activated on a fill of a line from higher level caches, is untouched, adding no extra gate depth. \par \underline{Placement disruptions}: There are no additional routing hotspots, congestion leading to placement perturbations, achieved by careful placement of added near memory logic. 
 Most near-memory logic is placed near the fill datapath, which isn't timing critical due to low logic depth, allowing it to tolerate wire delay. Only final stage is near read-datapath, and its low logic depth doesn't disrupt pipelines. Therefore, there's no performance/timing impact from added logic, and SPARK incurs no dynamic power cost as it can be fully gated.
\begin{figure}[t]
\centering
\includegraphics[width=\linewidth]{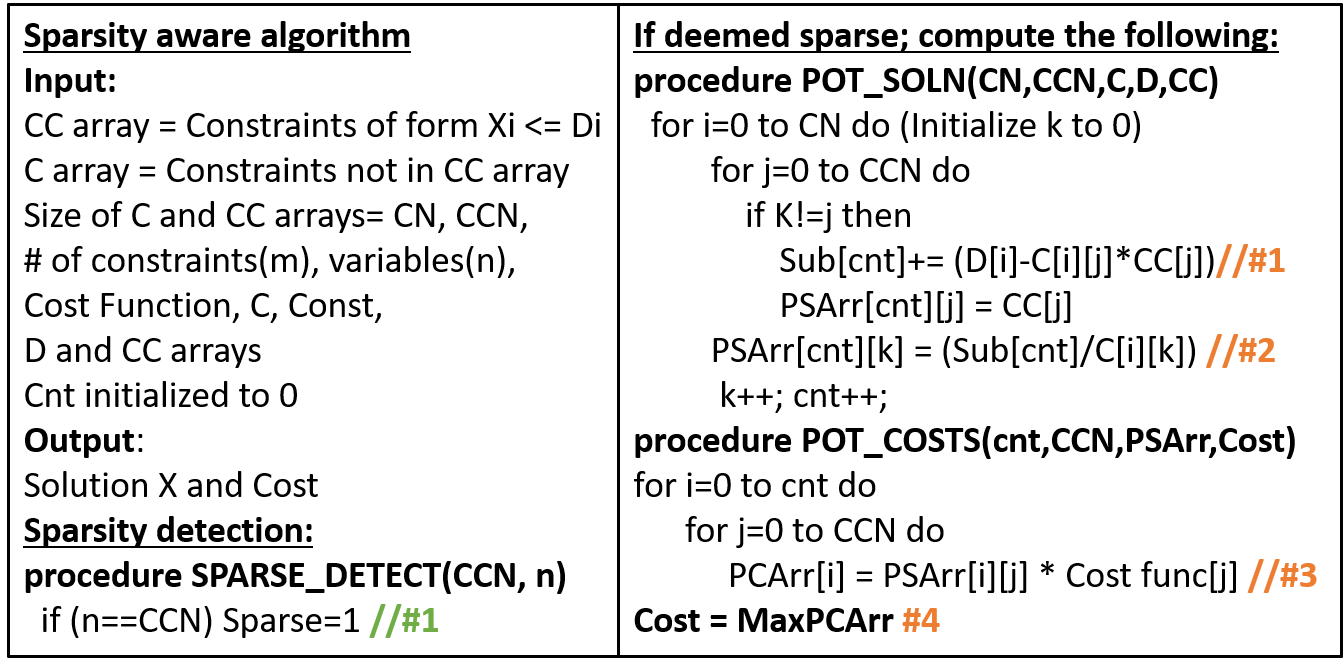}
\vspace{-2em}
\caption{ \textbf{\underline{Proposed sparsity-aware (SA)} algorithm begins with the detection of sparsity in FC engine. If deemed sparse, SA engine executes the proposed SA algorithm, by identifying potential solutions, and costs. }} 
\label{SA_algorithm}
\vspace{-1em}
\end{figure}
\par \underline{Latency, access ports, energy, capacity, associativity}: SRAM latency remains unchanged, as no additional logic is added to the read/write datapath. SPARK's near-memory logic only engages after the computed output from the memory array is captured in the row-buffers (flip-flops).
\par At the bitcell level, there is no change in \underline{access ports}. We reuse the decoupled read port in high-performance 8T SRAM for compute, maintaining  1 read, 1 write port. At the array level, the number of compute accesses matches that of simultaneous read accesses, since we reuse the existing row/column-decoding logic, leaving the access ports unchanged.
\par \underline{Read/write energy} remains unaffected, as the memory array's read/write datapath isn't altered for compute. The only additional power comes from precharge, due to 2:1 multiplexer for selecting between Vcc/2 and Vcc, adding just 0.001pJ, since it's shared across a column. Associativity, capacity and hit/miss detection circuitry remains unaffected. Like modern processors, we use way-predictor to identify way to access, firing the tag array to confirm correctness. In compute mode, data is "computed" rather than just array read.
\par \underline{Cache Coherence}: Among C*,D*,X* stored in memory, cache lines containing C,D do not undergo update, while X undergoes update. Incase lines containing C*/D*/X* get replaced or X* get updated, coherence in traditional CPUs is reused for communicating to other cores. We assume MESI protocol. Memory consistency is unaffected, as existing ordering requirements between memory operations is unaltered.

\section{SPARK's FULL-STACK APPROACH}

\subsection{Sparsity-aware algorithm} 
We propose a sparsity-aware algorithm (Fig.~\ref{SA_algorithm}), explained both mathematically and graphically. \par \underline{Mathematical understanding}: The algorithm starts by detecting sparsity (\underline{SPARSE\_DETECT}) in ILP problems.
In an ILP problem with m constraints and n variables (\textit{m $\geq$ n}), the algorithm classifies constraints as either cardinality constrained (CC) or general. Specifically, constraints of the form \textit{X\textsubscript{i} $\leq$ D\textsubscript{i}} are added to the CC array, while other constraints are placed in the general constraints (C) array. When the CC array contains exactly "n" elements, the ILP is considered sparse, indicating that there is a reduced number of active constraints relative to the total possible number. Following sparsity detection, the algorithm proceeds to the POT-SOLN function, which identifies potential solutions. This is achieved through efficient dot-product operations performed between the C and CC arrays. After the dot products are computed, subtraction and division steps are applied to further identify potential solutions. Finally, the POT-COSTS function (\#3) is responsible for determining the maximum and minimum cost values associated with the identified potential solutions.

\par \underline{Graphical understanding}: In an n-dimensional space, CC array elements form parallel planes, while C array elements form non-parallel planes. The intersection of these planes gives the optimal solution. Substituting values from n-1 CC array elements into n-1 C array variables yields the n\textsuperscript{th} variable for all constraints (\#1, \#2 in POT\_SOLN). X vectors represent potential solutions, and their costs are potential costs.

\par Dataflow initiates with L1 cache reading C/D/cost function from DRAM/L2, followed by sparsity detection in FC engine. For sparse ILP, SA engine executes the proposed sparsity-aware algorithm; for dense ILP, the Jacobi/reuse-aware B\&B approach runs in SLE/B\&B engines (shown in Fig.\ref{Architecture_ILP}a). 

\begin{figure}[t]
\centering
\includegraphics[width=\linewidth]{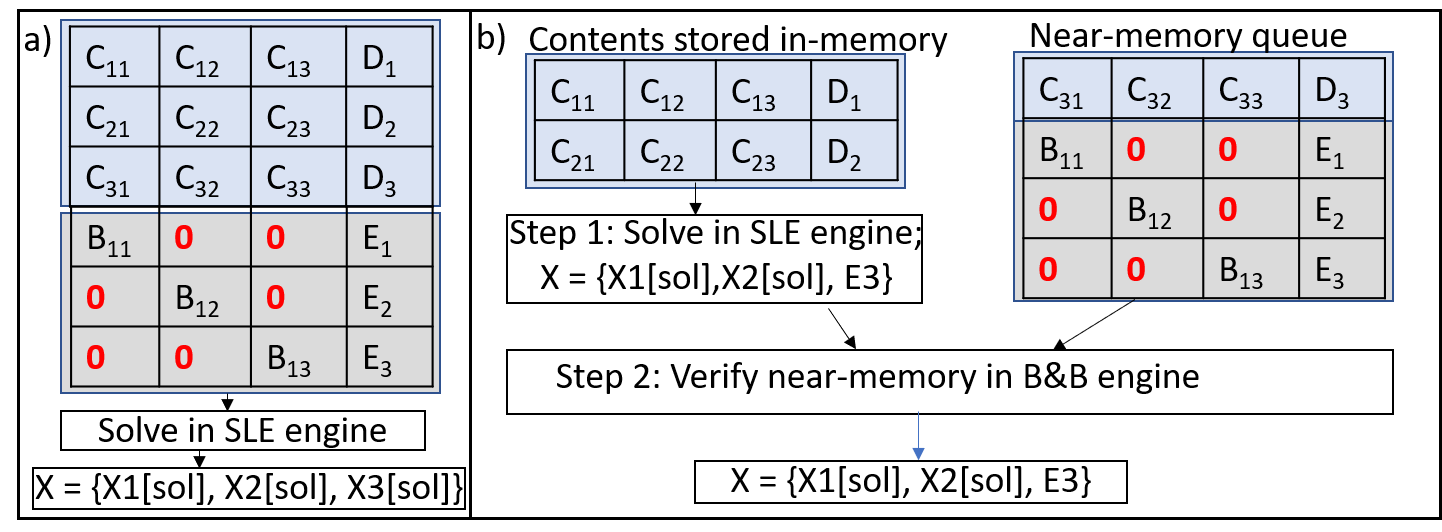}
\vspace{-2em}
\caption{\textbf{\underline{Reuse-aware B\&B} - a) B\&B adds sparse constraints (gray) to originally dense (blue) ILPs, and is solved by reusing SLE engine for B\&B without dedicated B\&B hardware, but suffers from energy-inefficiency. b) Proposed approach overcomes this by having near-memory queues}}
\label{Reuse_aware_BB}
\vspace{0em}
\end{figure}

\subsection{Reuse-aware B\&B algorithm for low area}
\par \underline{Idea}: To reduce area for B\&B acceleration, we propose a reuse-aware approach that allows hardware sharing between B\&B and SLE. SPARK reuses the SLE engine for B\&B due to the similarities in their computations.

\par \underline{Observation}: With each branch, an ILP with n constraints expands to "n+m" constraints, where "m" is the branching tree depth. For an ILP with 3 dense and sparse constraints after 3 B\&B levels (n=3), there are two options: i) Reusing the SLE engine for B\&B without area overhead, or ii) Adding hardware with a reuse-aware approach.

\par \underline{Tradeoff analysis}: The first option (Fig.\ref{Reuse_aware_BB}a) causes energy inefficiency and SLE under-utilization by solving additional sparse constraints in the SLE engine. The second option (Fig.\ref{Reuse_aware_BB}b) improves efficiency by solving only the first 2 dense constraints in SLE, while verifying the remaining 4 constraints near-memory, using parallel logic for 3 sparse constraints and MAC for 1 dense constraint. This boosts energy efficiency and compute density in the L1 cache. The near-memory queue for sparse elements further improves energy efficiency by 30\% and compute density by 20\%.
 \par 

\begin{figure}[t]
\centering
\includegraphics[width=\linewidth]{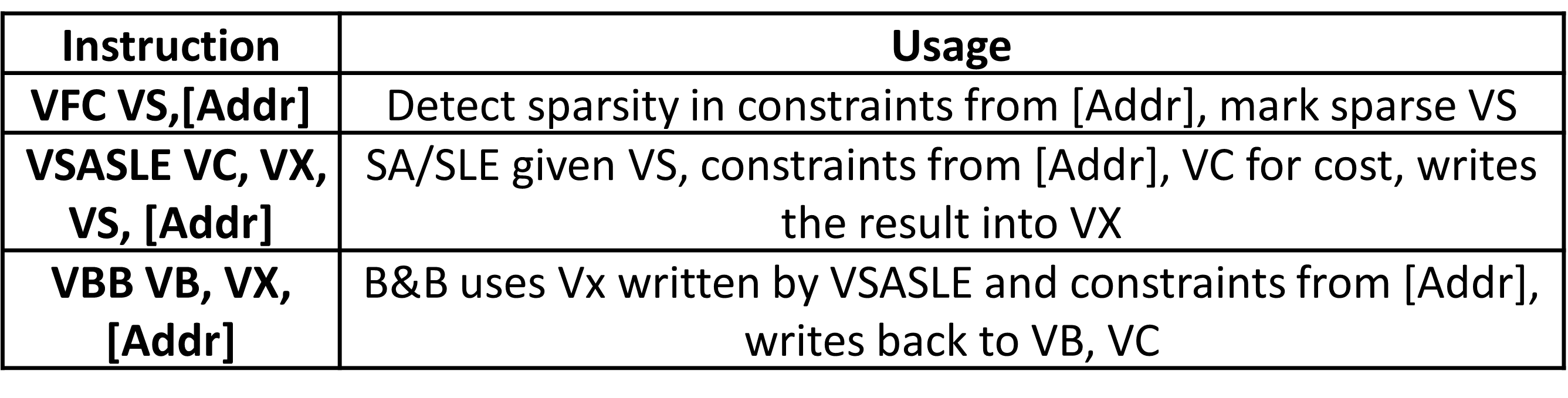}
\vspace{-2em}
\caption{\textbf{SPARK's additional instructions}}
\label{ISA_ILP}
\vspace{0em}
\end{figure}
\begin{figure}[t]
\centering
\includegraphics[width=\linewidth]{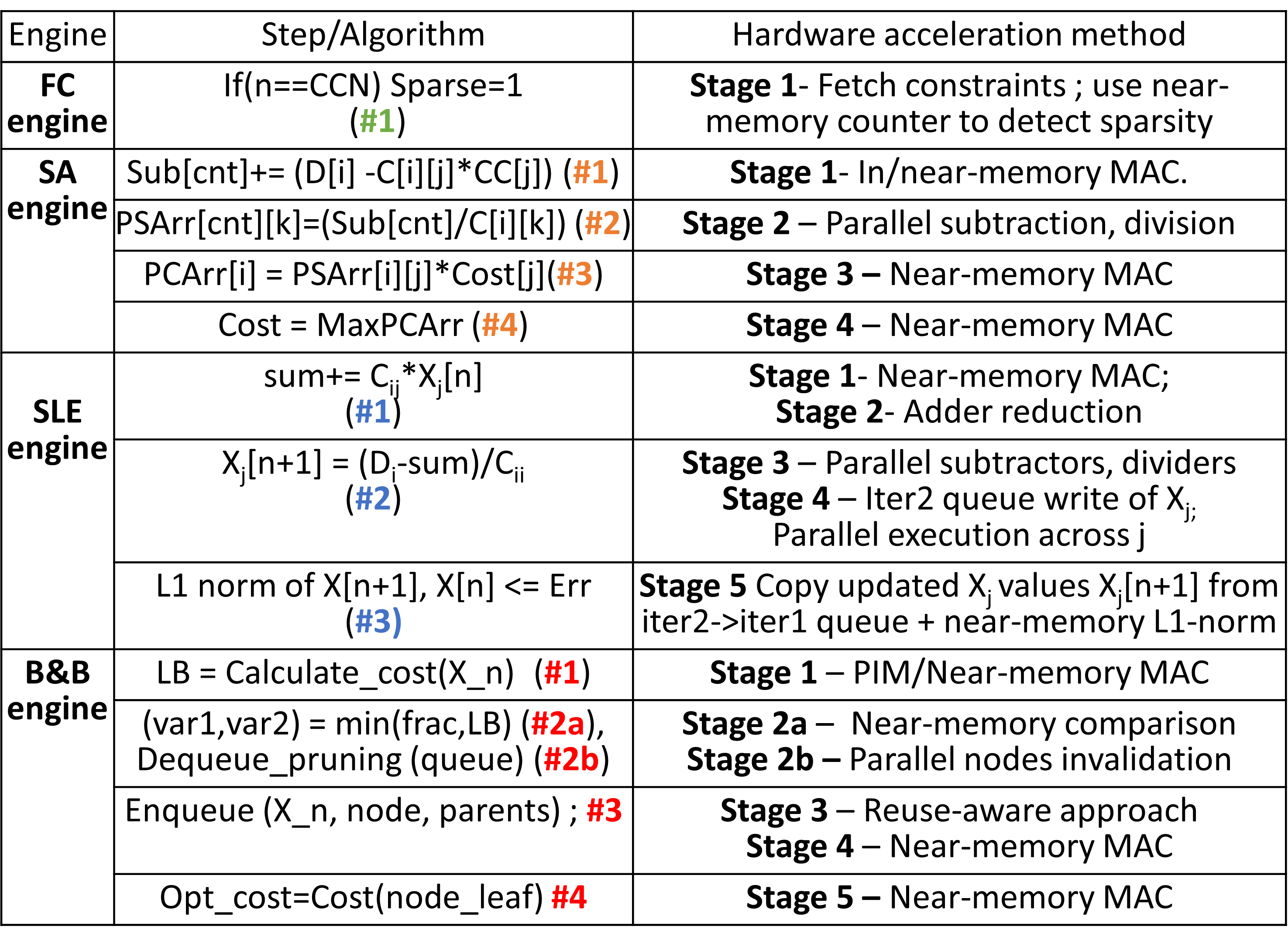}
\vspace{-2em}
\caption{\textbf{ Acceleration strategy for algorithms in Fig.\ref{SLE_BB_algo}, Fig.\ref{SA_algorithm}}}
\label{Overall_architecture}
\vspace{-1em}
\end{figure}
\subsection{SPARK's programming model}
Unlike dedicated accelerators, which often require the use of specialized and complex programming models, SPARK offers a more seamless integration by leveraging existing programming models such as sequential, multithreading, parallel, functional, and others. This flexibility is made possible because SPARK reuses the CPU microarchitecture with minimal additional instructions, meaning it can operate within the frameworks developers are already familiar with. Moreover, the modifications introduced by SPARK occur primarily at the compiler level, meaning that the underlying changes in the system are largely transparent to the programmer. This design choice ensures that the impact on the programming model is minimal, allowing SPARK to work effectively with any existing CPU programming model. 

 \subsection{SPARK's ISA modifications}
 SPARK introduces a set of new instructions and registers that significantly enhance the CPU's capabilities in handling specialized tasks. A control register is added that can be programmed to configure L1 cache to compute mode. This can be achieved similar to writing system registers (eg. MSR in ARM). If needed, the system can easily reset this register to return to a non-compute mode, ensuring flexibility and control over the processor’s operating state. When in compute mode, new instructions such as VFC, VSASLE, and VBB (Fig.\ref{ISA_ILP}) are decoded in front-end, signaling the back-end of the core for performing high-throughput near-memory execution \cite{Tensi}. 
 Furthermore, another alternative is to perform this compute using existing CPU instructions : CPU instructions are used for programming different engines in Spark. An additional VLD-PIM instruction is added to indicate that the L1 cache is used for PIM. Since execution on SA/SLE engines is a micro-architectural modification for sparsity awareness, software remains incognizant of the underlying sparsity-aware execution. Thus, both engines can be programmed using the following instructions with V indicating vectorized. The data is fetched from the L1 cache using VLD (vector load instruction) followed by the instruction sequence:In the case of using Jacobi iterative method for SLE solving, the rows of C need to be written onto memory using a vector store (VSTR) instruction during every iteration. This is followed by an instruction sequence (named I\textsubscript{jacobi}) that repeats every iteration across all elements in X:
\begin{enumerate}

    \item VMLA (Multiply-Accumulate): In-memory Dot product of C and D matrices, followed by near-memory accumulation of computed values
    \item VSUB (Subtraction): Difference between MLA and C/D
    \item DIV (Division): Division of MLA result and C\textsubscript{ii}
    \item VSUB (Subtraction): Computation of the difference between X\textsubscript{j}[n+1] and X\textsubscript{j}[n] $\forall$ X\textsubscript{j}
    \item VADD (Addition): Calculation of L1 norm
    \item BR (Branch): Decision of continuing to the next iteration
\end{enumerate}

\par B\&B engine is programmed for B\&B compute to enable reuse-aware compute for every level in the tree as follows:
\begin{enumerate}
    \item CMP and FMAX (Compare and Floating point max-value): Branching node and variable identification using upper bound and maximum fractional values of X
    \item I\textsubscript{jacobi}: Calculate solution for every branching node
    \item MLA (Multiply-Accumulate): Computation of upper/lower bound in maximization/minimization problem
    \item CMP (Compare): Decision of termination of branching
\end{enumerate}
\subsection{SPARK's execution strategy}
\par \underline{VFC execution}: VFC directs the FC engine to detect sparsity, providing ILP sparsity status, crucial for step \#1 in Fig. \ref{SA_algorithm}. VFC starts by loading constraints into L1 cache. C, D, X vectors can span multiple cache lines without assumptions about their location. Each 64B cache line stores 32 16-bit coefficients. During compute, data is read using base address + offset, and the sparse bit in the VS register is set based on sparsity, repeating across constraints to assess overall sparsity.
 \par \underline{VSASLE execution}: If deemed sparse, VSASLE is executed in SA engine, else, executed in SLE engine. VX undergoes dot product compute with constraints stored in memory, with VX mapped onto L1 cache columns (Fig.\ref{SLE_engine_PIM}), for which small (0.001\% area overhead) decoding logic is added near-L1 cache. VX stores updated X values post each iteration, mapped onto iteration queues in hardware. During sparsity-aware compute, the address from memory points to either C or CC array elements (\#1-2 in Fig.\ref{SA_algorithm}), and points to constraints for performing Jacobi, in dense compute. SPARK's multiple cache banks help with achieving a high throughput of 32 16-bit MACs possible in a given cycle from a single core for a dense ILP.  
If a constraint crosses multiple CLs, we perform partial updates for each CL. VC is used for storing the initial cost and stores the updated cost, as we proceed through iterations. \par \underline{VBB execution}: The VBB instruction reads the contents of the VX register and, based on the data, activates the B\&B engine for the final ILP solution. This process follows the same principles as those used in the SLE engine, as SLE engine is reused for B\&B compute as well. If ILP is sparse, VBB acts as NOP, as B\&B engine can be gated during sparse compute. 
 \begin{figure}[t]
\centering
\includegraphics[width=\linewidth]{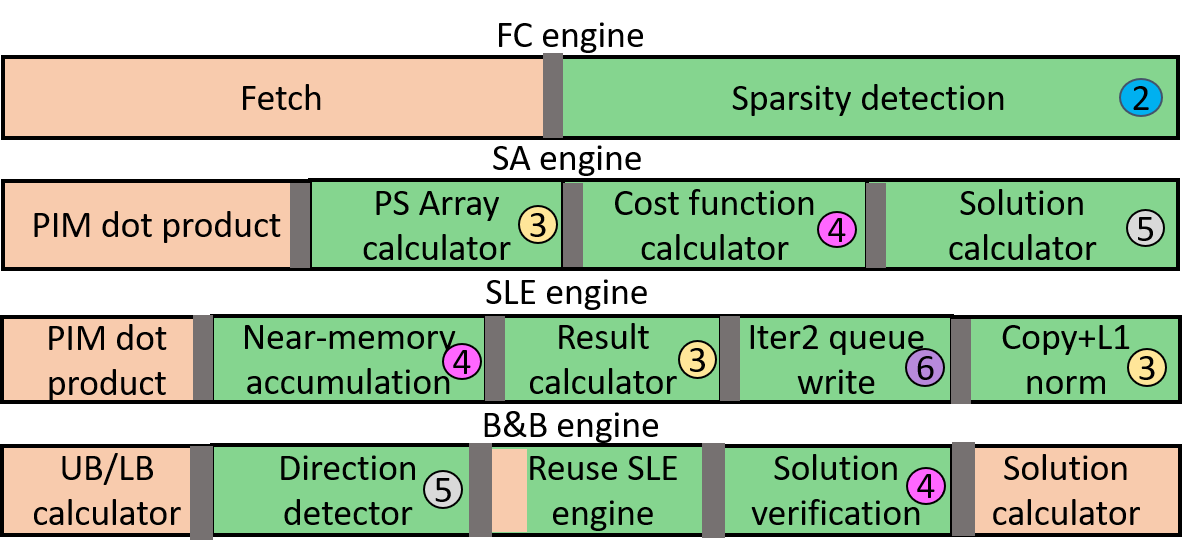}

\caption{\textbf{Spark's \underline{reuse-aware} pipelined architecture - (i) L1 cache (orange) used in all the engines for PIM compute  (ii) Reuse-aware B\&B approach enables usage of SLE engine for B\&B compute (iii) Resource sharing across engines (stages marked with same number). Gray boxes indicate pipeline registers. Numbers 2-6 indicate stages requiring additional hardware (green) }}
\label{CPU_GPU_proposed_2}

\end{figure}


\subsection{SPARK's execution flow for sparse ILP}
FC engine uses near-memory counters to detect sparsity. SA engine uses MAC, subtraction, division to solve sparsity-aware compute. SLE engine uses near-memory MAC, subtraction, and division, B\&B engine reuses SLE engine in step 3, near-memory MAC, subtraction, division, discussed in Fig.\ref{Overall_architecture}. 

 \begin{figure}[t]
\centering
\includegraphics[width=\linewidth]{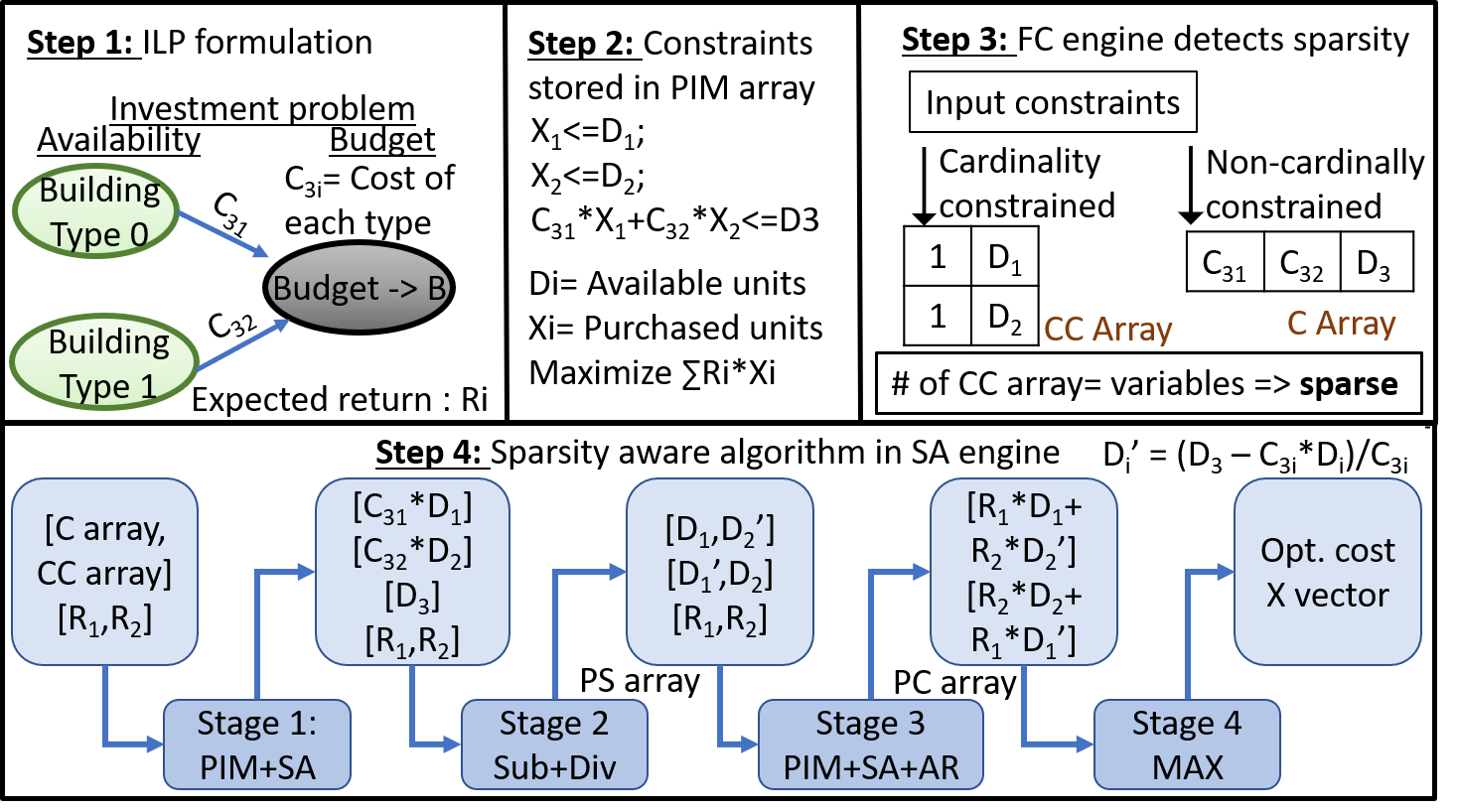}
\vspace{-2em}
\caption{\textbf{Step 1: Investment problem with sparse constraints is stored in L1 cache. Step 2: C matrix and D vector is fetched from the L1 cache in FC engine. Step 3: These are pushed onto either the CC or C array and is used for sparsity detection. Step 4: Sparsity-aware approach uses PIM's high throughput compute between C and CC array. }}
\label{SA_engine}
\vspace{-1em}
\end{figure}
\begin{figure}[t]
\centering
\includegraphics[width=\linewidth]{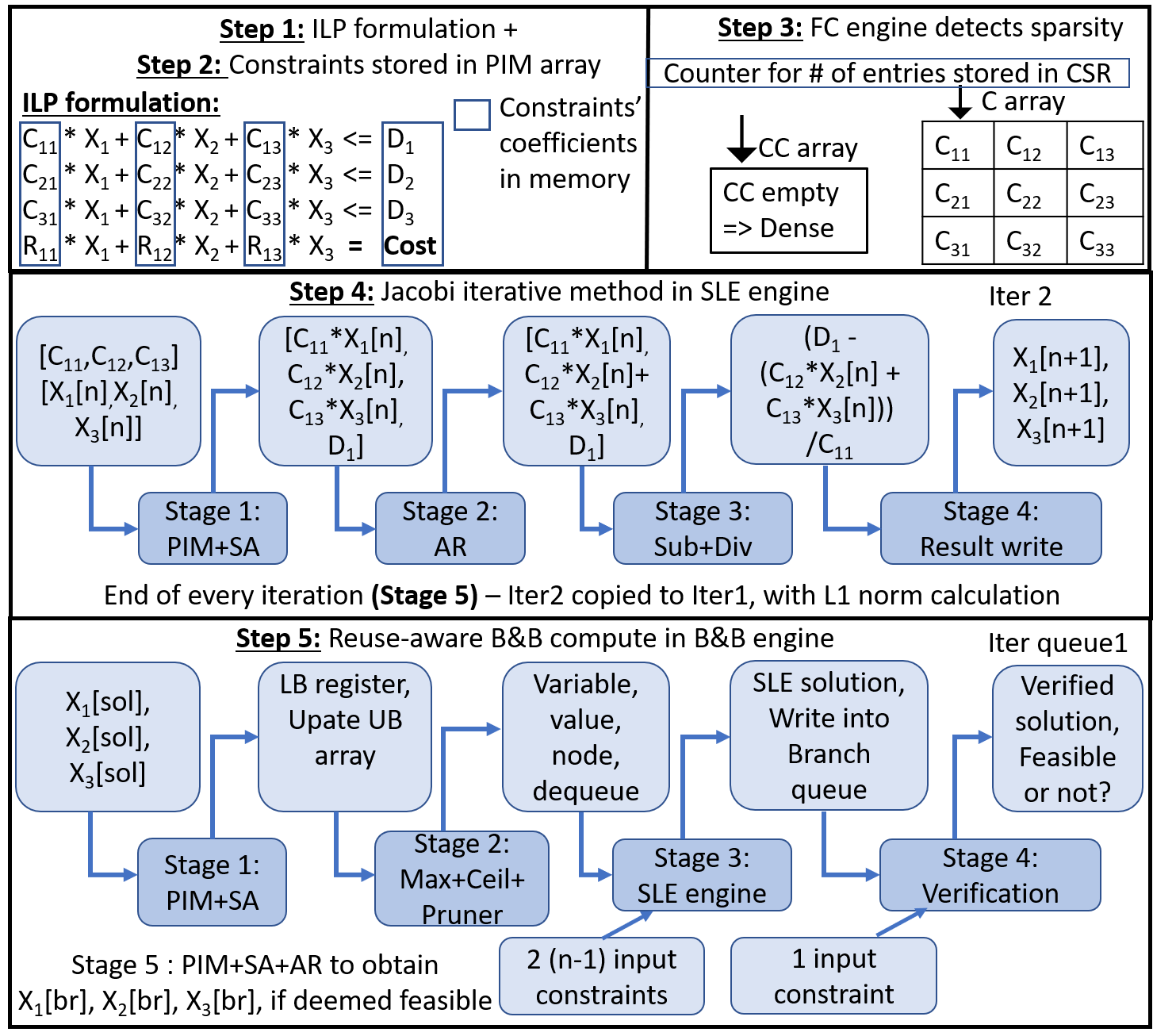}
\vspace{-2em}
\caption{\textbf{ ILP with 3 constraints (for example) is stored in the L1 cache in Step 1. In step 2, C and D matrices in the L1 cache are read out and in step 3, the FC engine detects the problem to be dense, as CC array is empty. Jacobi iterative method is executed in Step 4 and the reuse-aware B\&B approach in B\&B engine accelerates B\&B in step 5. }}  
\label{SLE_BB_engine}
\vspace{-1em}
\end{figure}
\par \underline{Step 1} shows ILP formulation for investment problem by using FC and SA engine. \par \underline{Step 2} shows PIM (L1 cache) array contents, for illustration Fig.~\ref{PIM}.a. If there are \textit{n} constraints of the form \textit{X\textsubscript{i}$\leq$D\textsubscript{i}} and \textit{(m-n)} constraints of the form \textit{$\Sigma$C\textsubscript{i}*X\textsubscript{i}$\leq$D\textsubscript{i}}, n constraints of the form are stored first, followed by \textit{m-n} constraints, to detect sparsity early and turn off SLE, B\&B engines. Fig.~\ref{SA_engine} illustrates this by storing X\textsubscript{1-2}$\leq$D\textsubscript{1-2} followed by C\textsubscript{31}*X\textsubscript{1}+C\textsubscript{32}*X\textsubscript{2}$\leq$D\textsubscript{3} in Step 2. 
\par \underline{Step 3} checks if a constraint is sparse by counting its non-zero coefficients. If there are 2, it goes to the CC array; otherwise, it goes to the C array. The first two constraints store X\textsubscript{i} and D\textsubscript{i} in the CC array, while the 3\textsuperscript{rd} constraint is pushed to the C array. Sparse constraints are counted in the CC array to determine if the ILP is sparse. The FC engine filters out zero coefficients from C/CC arrays.
\par \underline{Step 4} performs sparsity-aware ILP compute in SA engine using PIM's high throughput. Stage 1 executes a near-memory MAC between elements in C (stored in memory) and CC arrays (mapped to column). Stage 2 finds potential solutions (PS) using near-memory subtractors/dividers. In stage 3, cost of each PS is computed using near-memory MAC and enqueued into the PC array, with optimal cost found in stage 4.

\subsection{Execution flow for dense ILP}
\par\underline{Step 1}: Fig.~\ref{SLE_BB_engine} shows ILP with 3 dense constraints stored in L1 cache. 
\par \underline{Step 2,3}: Constraints are read from L1 cache, and the ILP is deemed dense since the CC array is empty, based on the approach mentioned for sparse ILPs.
 \par \underline{Step 4}: ILP is executed using SLE.  
Stages 1-2 in SLE engine execute near-memory MAC (Fig.~\ref{PIM}c) with stage 3's divider computing step \#2 in Fig. \ref{Overall_architecture}. Stage 4 updates the result into the Iter2 queue. In stage 5, L1 norm compute determines whether the problem has converged. The final solution (X\textsubscript{1-3}[sol]) is transferred to B\&B engine. 
 \par 
\par \underline{Step 5}: B\&B compute for the initial branching tree level is shown using X\textsubscript{1-3}[sol], assuming the final solution (X\textsubscript{1-3}[br]) is achieved after one level of branching. In (i) Stage 1, global LB/local UB is calculated by reusing PIM for dot-product and storing in global LB register/local UB array. The local UB array/queue is enqueued after each branching tree level, while global UB remains constant (ii) Stage 2 identifies branching nodes, values, variables, and decisions for parallel pruning using max/ceil/max functions. (iii) Stage 3 capitalizes on parallelism from the SLE engine. (iv) Stage 4 uses reuse-aware approach to verify the solution, by a) reusing PIM, b) using simple MAC without replicating SLE engine. 3 arrays store branching values, variables, and indices of parent nodes, for cases where child node is invalidated along with parent node.  (v) In Stage 5, the final solution is obtained through MAC.

\subsection{Execution flow for dense/sparse LPs} Sparse/dense LPs use same flow as their ILP counterparts. In LPs the final solution comes from SLE, as B\&B is unused. 
Revisiting SLE algorithm in Fig.~\ref{SLE_BB_algo}, using integers for C/D in-memory enables add/sub/mul operations with the mantissa, given identical exponents across X. Steps \#1,\#2 are resolved to an integer dot-product/subtraction between mantissa of X and C, with divider in step \#2. Steps \#1, \#2 are repeated till convergence.

 \section{EVALUATION METHODOLOGY}
\subsection{Benchmarks}
MIPLIB 2017/2010 \cite{MIPLIB_2017} consists of real-life ILPs developed to analyze the performance of different ILP solvers. 
 We chose 7 benchmarks (a mix of L1-cache fitting and overflowing benchmarks) to evaluate SPARK's benefits. For instance, NS, ST, BL do not fit inside the L1 cache, to study SPARK's effectiveness for large-sized workloads. Some of the common problems described are:
\textbf{Transportation problem}: ILP problem with fairly dense constraints, in which products are transported from "m" sources to "n" destinations, subject to demand and supply constraints, in a way that minimizes the total transportation cost. The supply/demand constraints are such that the \# of units transported from i\textsuperscript{th} source to j\textsuperscript{th} destination should be less/greater than the supply from i\textsuperscript{th} source/demand for j\textsuperscript{th} destination. A row in C and D matrix is formed by the \# of transported units and demand/supply units, respectively. This makes use of FC, SLE and B\&B engines.
\par \textbf{Real Estate investment problem}: ILP problem with sparse constraints making use of FC and SA engine, maximizes the income given the cost of a building and the \# of buildings that one can buy subject to a constraint on the investment budget. A row in C/D matrix in ILP formulation consists of the \# of buildings and the budget, respectively.

 \par 
 
 \subsection{Simulation methodology}
 SPARK is compared to software-optimized ILP execution on multi-core CPU/GPU, as there are no prior ILP accelerators, and traditional accelerators fall short of CPU/GPU performance (Fig.\ref{Comparison_new_accelerators}). SPARK's performance is modeled using a C++-based cycle-accurate simulator with a prefetching strategy to hide data movement latency, and it simulates PIM array intricacies at the RBL/RWL level, which existing CPU simulators (Gem5) cannot. Python API-based ILP execution provides end-to-end application performance, including setup and data loading effects seen in CPU/GPU execution. \subsection{SPARK model} SPARK's model includes 32KB I/D cache, 64B cache line size, 4MB shared L2 cache, LRU replacement, 2GB DRAM, 5-wide decode, 8-fetch width, 32-entry load/store queue, and a stride-2 prefetcher for the PIM-capable L1.
To validate SPARK's CPU modeling (without PIM mode), we compare to Gem5 using 16 SPEC benchmarks (8 INT/8 FP), covering L1 cache hits (eg. sjeng, hmmer) and misses (eg. bzip2, gcc, gobmk) to evaluate prefetching/memory performance. SPARK’s metrics, including L1 cache hits, miss latency, execution time, align within 0.2\% of Gem5 results.

\subsection{SPARK micro-architecture} 8T SRAM-based L1 cache array is organized into 16 banks, each with 256 rows and 256 columns, optimized for PIM compute. The near-memory logic includes:
a) A 32-bit counter in the control stage's cardinality checker for sparsity detection.
b) The SLE engine includes two 256-entry centralized arrays repurposed for potential solutions and cost arrays in the SA engine.
c) The B\&B engine has a shared Global UB/LB register with 1024 entries, including UB/LB and branch variable/value arrays.
d) Subtractors/dividers are set at 1 per bank, with adders at 1 per 16 columns.
Energy for compute/read is based on RBL discharge when RWL is ON, with 40fF/35fF capacitance at 1V. SRAM latency is 2ns, and data movement costs 1pJ/bit \cite{Horowitz}. We describe near-memory logic in System Verilog and synthesize SPARK's digital components for area, power, and energy estimates using Synopsys Design Compiler with 45nm FreePDK technology \cite{45nmPDK}, operating at 1V with a 2ns clock.
 \subsection{CPU/GPU comparison} Multi-core CPU is AMD's Zen3, using ILP solver Gurobi, multi-threading/cores and AVX. 
 The GPU used is NVIDIA's Tesla v100, along with cuSparse to solve sparse ILPs. 
 Execution times are recorded for performance comparison. Power for GPU is measured using Nvidia System Management Interface (nvidia-smi). The power in the idle state is deducted from the power associated while executing an ILP to separate power usage from other processes. In CPUs, power is measured using power-stat, discounting idle power. Energy is obtained by multiplying time with power.  
       
\begin{figure}[t]
\centering
\includegraphics[width=\linewidth]{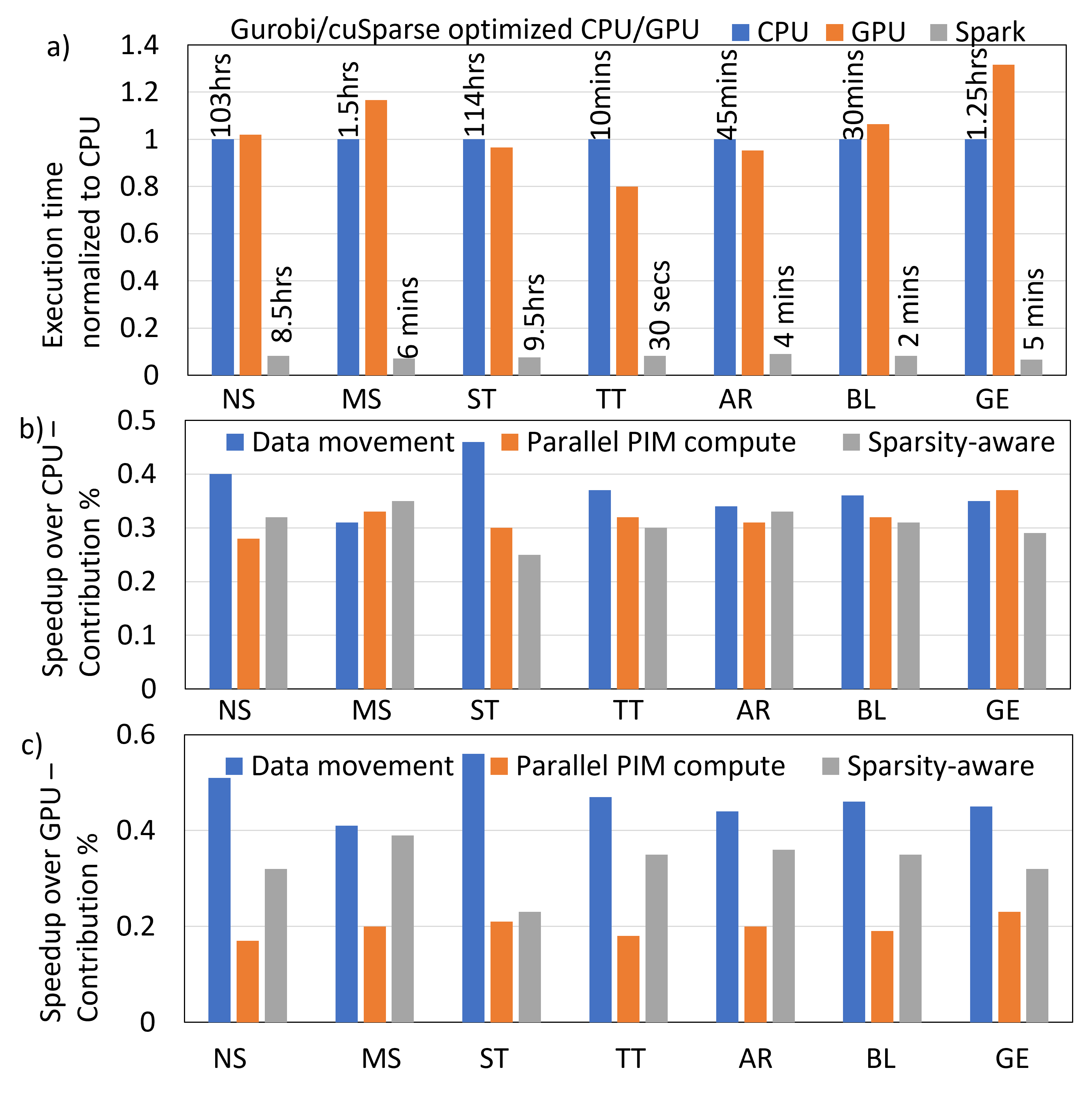}
\vspace{-2em}
\caption{\textbf{ \underline{Speedup of Spark for sparse ILP}: a) Spark shows 12-15x/12-20x speedup over Gurobi/cuSparse optimized CPU/GPU.{ Relative contribution of reduced data movement, parallel compute, and sparsity-aware compute for improvement over b) CPU c) GPU.}}}
\label{Perf_ILP}
\vspace{-1em}
\end{figure}

 \subsection{Performance Breakdown Evaluation}
 SPARK's benefits come from i) reduced data movement due to in/near-memory compute alongwith prefetching. ii) high throughput of parallel PIM compute. iii) Sparsity-awareness. We identify their relative contributions: For iii), we get rid of sparse datapath and compute using dense datapath. For ii), we model PIM's throughput to be 1 op/cycle to mimic non-parallel PIM compute, while the remaining benefits come from the reduced data movement aspect.

\section{RESULTS}
\subsection{Performance comparison for sparse ILP wrt CPU}
 Fig.~\ref{Perf_ILP}.a shows the comparison of execution time measured using Spark and software-optimized multi-core CPU (AMD Zen3)/GPU ILP, running MIPLIB benchmarks, considering application-level tradeoffs.  The execution times in CPUs are at least 12x-15x higher than in Spark due to reduced data movement, increased throughput in PIM and early detection of sparsity with sparsity-aware execution leading to reduced number of insignificant computations to complete iterations faster. This is valuable in cases where the sparsity is less (70-80\%) like in MS, AR. Fig.~\ref{Perf_ILP}.b shows that data movement cost in large workloads is higher, parallel PIM compute is useful uniformly across all workloads, sparsity-aware compute is more valuable in highly sparse workloads. Speedup is higher for workloads that fit in L1, due to reduced data movement 
 \subsection{Performance comparison for sparse ILP wrt GPU} Despite cuSparse optimization, GPU performance lags behind the CPU, while SPARK achieves a 12–20x speedup over the GPU due to (i) the absence of host-GPU interaction overhead, as SPARK integrates seamlessly into the CPU pipeline. (ii) The data transfer overhead for dot product computation is reduced by performing in/near-memory compute. (iii) The near-memory sparsity-aware algorithm minimizes hardware underutilization from sparsity/B\&B by performing only useful compute with PIM, as seen in NS, ST, TT, AR, and BL, where sparsity is very high.
 (iv) cuSparse is ineffective for MS, GE than Gurobi, where sparsity is low, resulting in longer execution time than CPU, achieving 20x speedups. Fig.~\ref{Perf_ILP}.c shows similar trends as that of Fig.~\ref{Perf_ILP}.b.

  \begin{figure}[t]
\centering
\includegraphics[width=\linewidth]{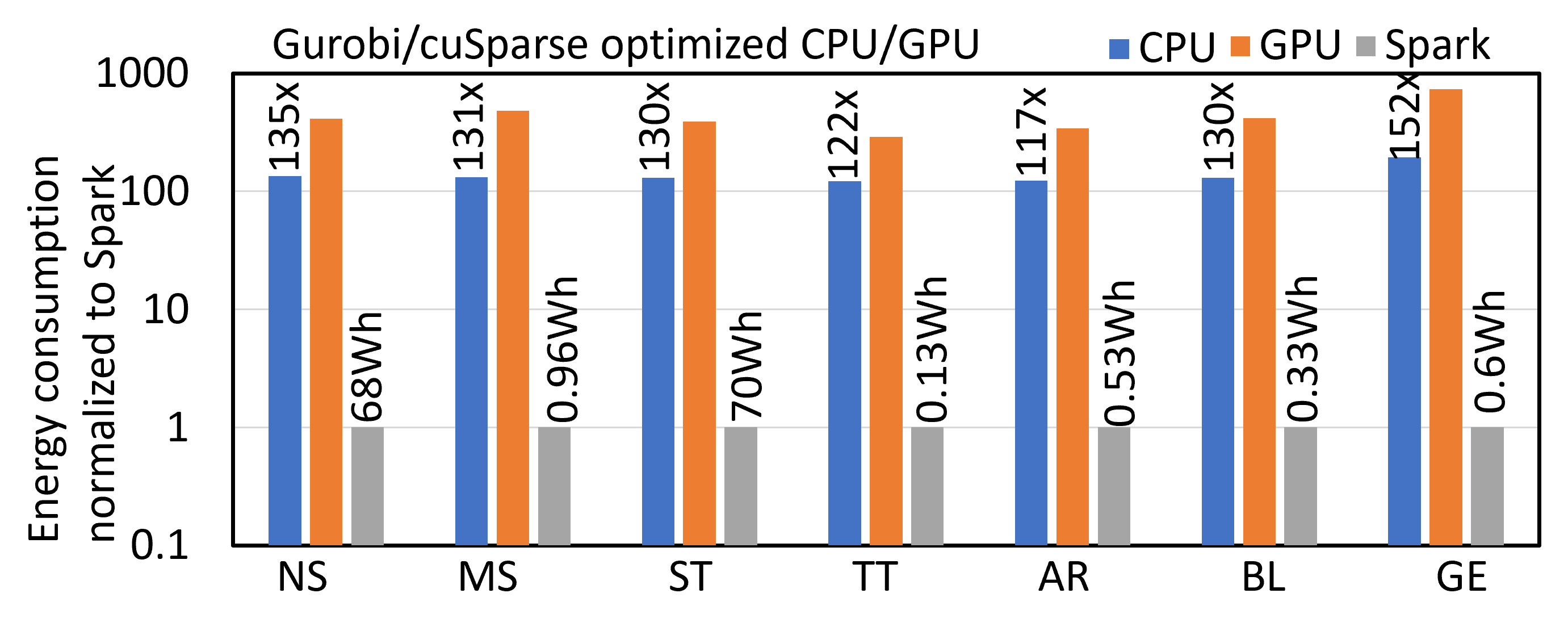}
\vspace{-2em}
\caption{\textbf{ Spark shows 117-152x/400-740x improvement in energy for sparse ILP over CPU/GPU. Note: y-axis uses log scale.}}
\label{Energy_ILP}
\vspace{-1em}
\end{figure}

\subsection{Energy comparison for sparse ILP wrt CPU}
 Spark shows 117-152x energy improvement over CPU, considering power, execution time (Fig.~\ref{Energy_ILP}). For CPU and Spark, the average power required is approximately 80-90W for the CPU and 7-10W for Spark. This substantial difference in power consumption is primarily due to Spark’s design, which significantly reduces data movement, incorporates early sparsity detection, and utilizes a reuse-aware architecture. These optimizations help minimize power consumption in Spark compared to the CPU. In addition to offering significant energy improvements of 120x in extremely sparse workloads, Spark achieves even higher energy savings of 150x in GE, a less sparse workload.  This improvement is driven by early sparsity detection, which enables the system to shut off unused engines, further reducing energy usage. 
 \subsection{Energy comparison for sparse ILP wrt GPU}
 We observe 400-740x energy improvements over the GPU (which averages $~$250W). Specifically, the GPU's streaming engine is often underutilized, and there is frequent host-GPU data movement, both of which contribute to inefficiencies in energy usage. Additionally, GPUs lack the specialized execution units required for optimal energy efficiency in certain workloads. These issues are effectively mitigated by leveraging near-memory sparsity and reuse-aware compute strategies, which optimize the computation process directly within memory, reducing unnecessary data movement. For less-sparse workloads like GE, the energy improvements are higher (740x) because of inefficient GPU compute.

\subsection{Performance/energy comparison for sparse LP}
We relax integer constraints from MIPLIB benchmarks, removing B\&B. GPUs struggle with sparsity and thread divergence. Fig.~\ref{Perf_LP}.a shows if CPUs outperform GPUs without B\&B divergence, the ILP problem is sparsity-bound (SB). Otherwise, it is divergence-bound (DB).
 In benchmarks like NS, speedups in solving LPs over CPUs are seen, but not significant, demonstrating the interaction between divergence and sparsity bounds, with divergence dominating sparsity. Spark shows a speedup of 7-20x/8-17x than CPU/GPU due to sparsity aware near-memory compute. For DB benchmarks, the enhanced performance of the GPU makes up for the overall power requirement. For SB benchmarks (MS, GE), CPU power overhead is higher. With SPARK’s near-memory sparsity-aware compute, the B\&B engine is turned off, yielding 103-272x/96-250x improvements over CPU/GPU (Fig.~\ref{Perf_LP}.b).

\par \subsection{Performance/energy comparison for dense ILP} We run randomly generated dense ILP constraints on CPUs with Gurobi and on GPUs without cuSparse. SPARK's near-memory SLE engine improves efficiency and performance. Fig.~\ref{Dense_ILP}a shows a 6-8x speedup over the CPU due to limited throughput and high execution latency. Thread divergence in B\&B reduces GPU throughput, leading to periodic host-GPU interactions, causing a 7-10x speedup for SPARK over the GPU. For dense ILP, we observe linear speedup for 1K-10K constraints, as convergence slows with more constraints. Fig.~\ref{Dense_ILP}b shows a 60-75x/180-210x energy improvement over CPU/GPU with reuse-aware near-memory approach.

\begin{figure}[t]
\centering
\includegraphics[width=\linewidth]{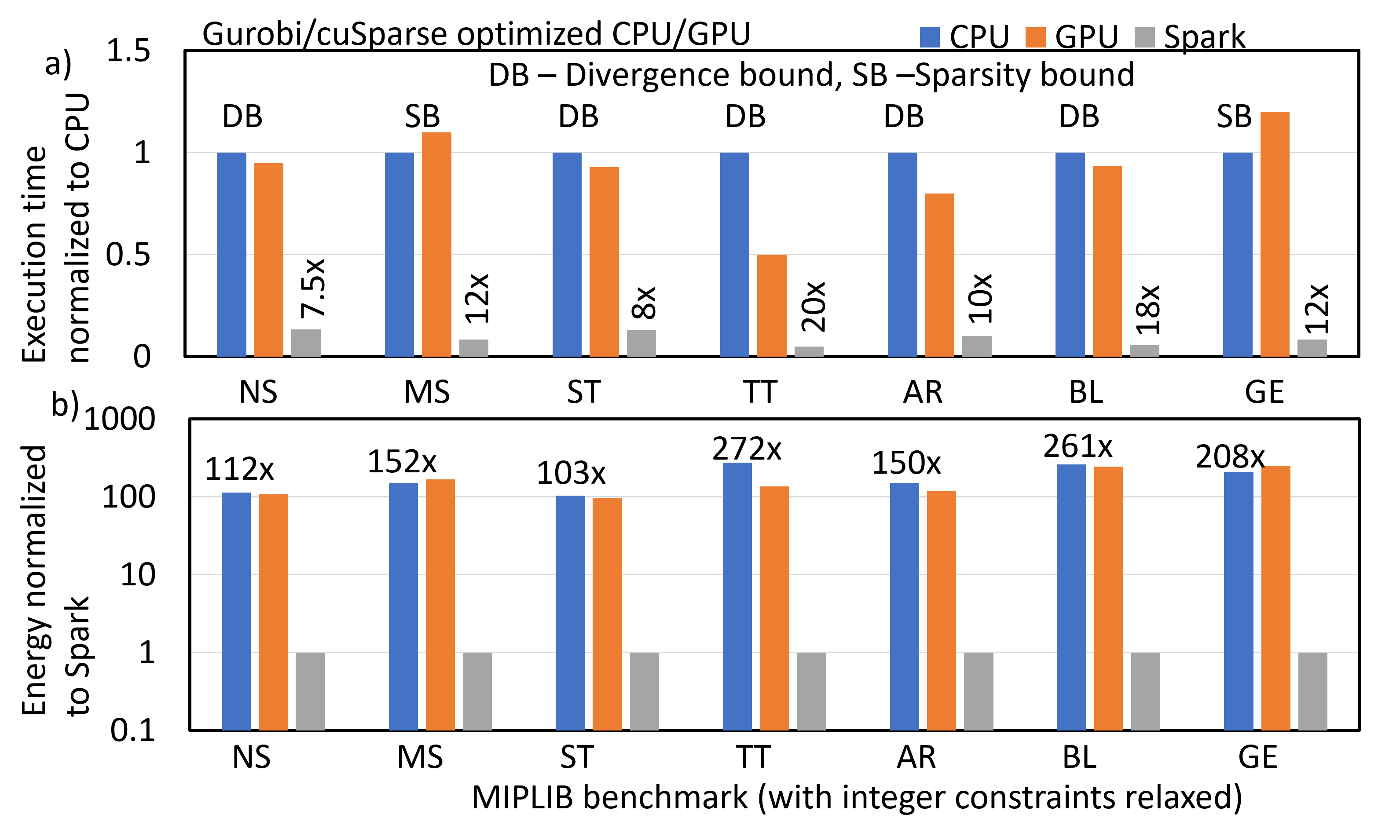}
\vspace{-2em}
\caption{\textbf{ \underline{Spark vs CPU/GPU for LP} - Sparse LP in SA engine- a) Performance b) Energy comparison between Gurobi (CPU) and cuSparse (GPU) decoupling sparsity (both SLE and B\&B) and thread divergence issues (B\&B) in GPU, as there is no B\&B overhead. Spark shows 7-20x/8-17x speedup/energy improvement of 103-272x/96-250x over CPU/GPU. }}
\label{Perf_LP}
\vspace{-1em}
\end{figure}

\subsection{Performance/energy comparison for dense LP}
 We relax the integer constraints of dense ILPs, and find the time/power for evaluating SLE, as there is no B\&B for LPs. 
 Fig.~\ref{Dense_ILP}.c shows a 4-5x speedup due to the near-memory high-throughput approach, while GPU requires frequent host-GPU interactions. Jacobi's speedup contribution ranges from 48\% to 51\% for 1K-50K constraints, with the rest from B\&B. The dataset's density increases GPU utilization, leading to better speedups than CPUs. Despite this, GPU power usage increases energy consumption, especially with Gurobi for CPUs up to 10K constraints. Beyond 50K constraints, energy decreases due to better streaming engine utilization. SPARK shows a 105-180x energy improvement over CPU/GPU by shutting off the B\&B engine and leveraging near-memory compute.

\subsection{SPARK's area analysis}
SPARK's area analysis is divided into three main components: i) memory array (bitcells), ii) peripheral circuitry (row/column decoders, multiplexers, and sensing circuitry), and iii) added near-memory logic (shifters and adders).
\par \underline{Memory array}: We use a 16-bank 8T SRAM array with a size of 0.08mm² per core. SPARK doesn't alter the memory array, as the existing RBL and contents are reused for in-memory computation, ensuring that the area occupied by the memory array remains unchanged.
\par \underline{Peripheral circuitry}: SPARK introduces a 2:1 multiplexer to enable efficient dot-product computation through RBL precharge. This results in increasing the area by 0.005mm² per core. Other multiplexers/decoders remain unchanged from the baseline peripheral circuitry.
\par \underline{Added near-memory}: SPARK requires 0.37mm² per core, including sparsity detection counters (0.03mm²/0.1\%), subtractors/dividers (0.11mm²/0.4\%), shift-add/adder (0.1mm²/0.4\%), comparators (0.02mm²/0.1\%), and control/queues (0.11mm²/0.4\%). Reduced area is due to resource sharing across SPARK engines.
\begin{figure}[t]
\centering
\includegraphics[width=\linewidth]{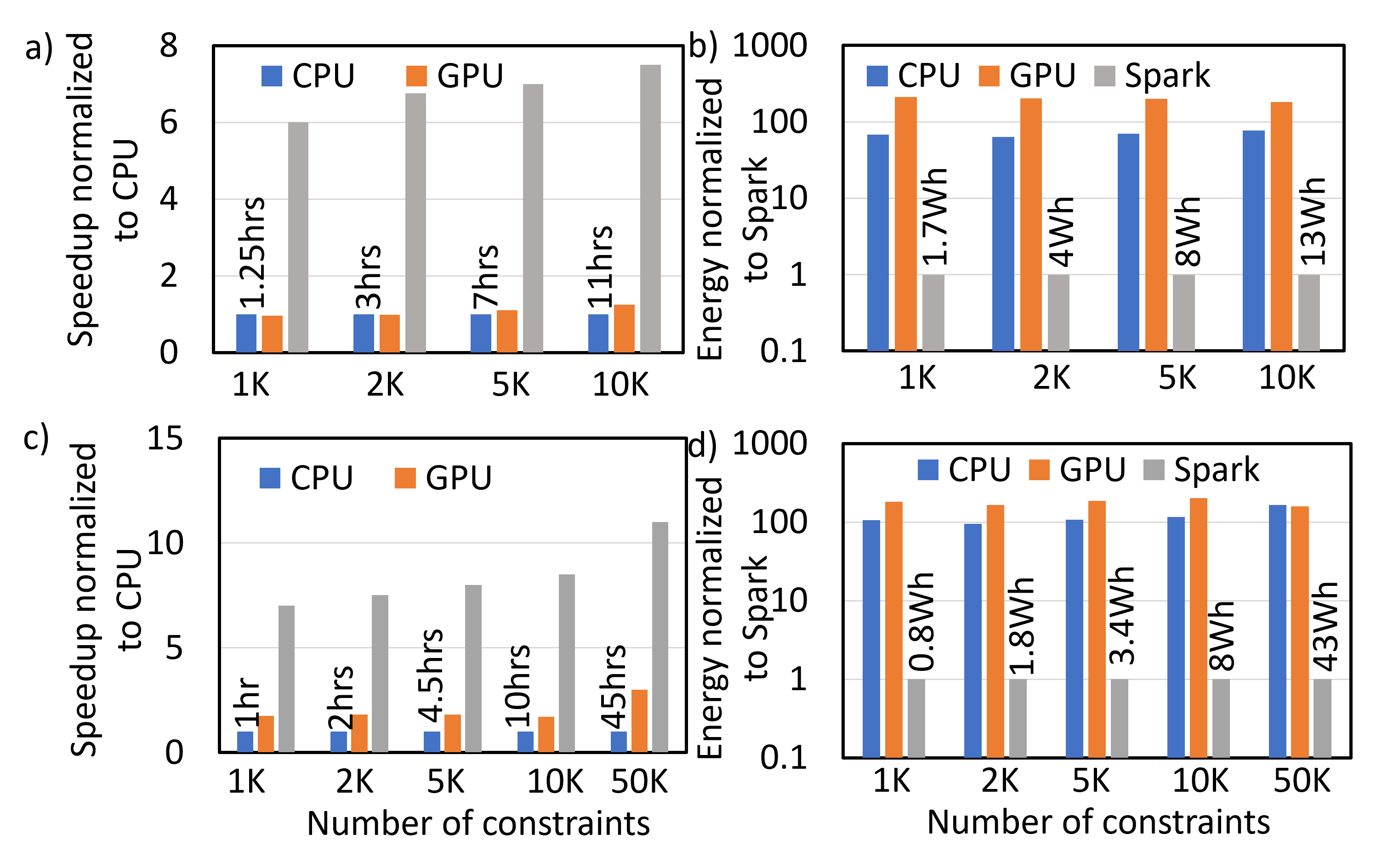}
\vspace{-2em}
\caption{\textbf{\textbf{a) \underline{Dense ILP speedup sensitivity to problem size }:} Speedup of 6-8x/7-10x
b) Energy improvement of 60-75x/180-210x over CPU+Gurobi/GPU \textbf{c) \underline{Dense LP sensitivity}: Speedup of 7-7.5x/4-5x over CPU/GPU d) Energy improvement of 105-115x/150-180x over CPU/GPU. }}}
\label{Dense_ILP}
\vspace{0em}
\end{figure}
\subsection{Comparison between Tesla A100 and V100}
We simulate using Tesla A100 in addition to the Tesla V100, with the results shown in Fig.~\ref{Tesla}. While the A100 boasts higher computational power and increased memory bandwidth compared to the V100, the observed performance gains are minimal and only become significant when processing large workloads. For smaller workloads, both GPUs show nearly identical performance in terms of solution time. This is because, despite the A100's higher throughput capabilities, the latency involved in data movement cannot be sufficiently offset by its increased bandwidth in smaller datasets. Furthermore, the A100 consumes more energy due to its higher power requirements, which may reduce its overall energy efficiency for tasks that do not fully leverage its enhanced capabilities, making V100 a power-efficient option for smaller workloads
 \begin{figure}[b]
\centering
\includegraphics[width=\linewidth]{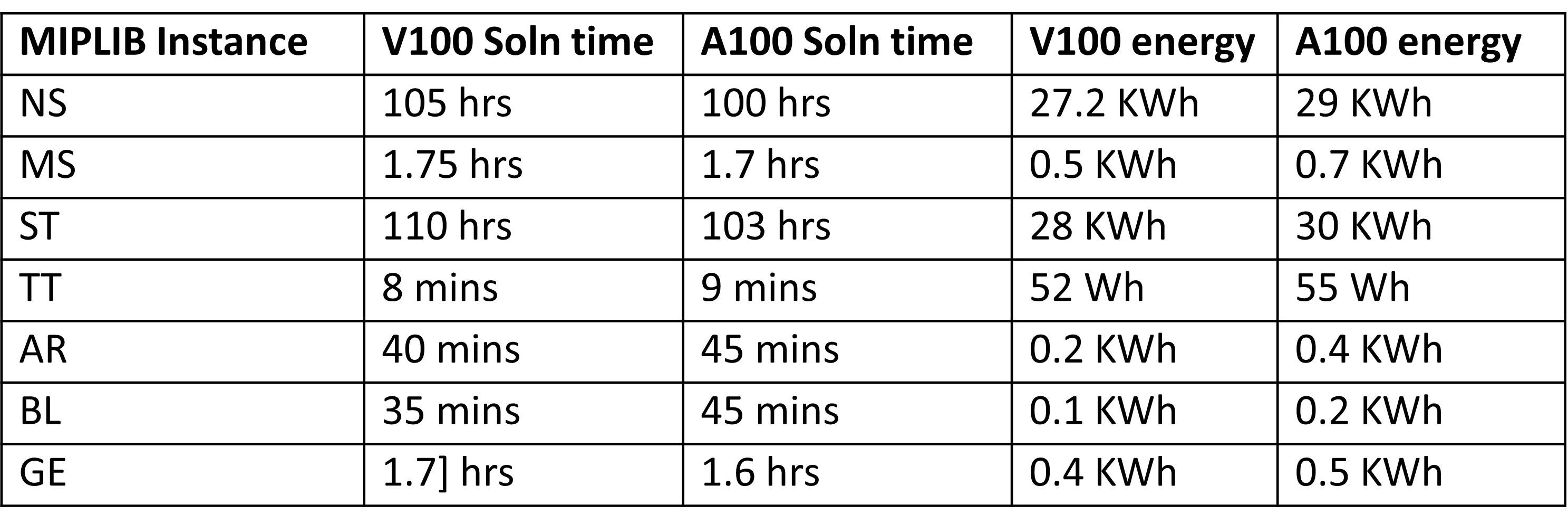}
\vspace{-2em}
\caption{\textbf{A100/V100 comparison }}
\label{Tesla}
\vspace{-1em}
\end{figure}

\begin{figure}[t]
\centering
\includegraphics[width=\linewidth]{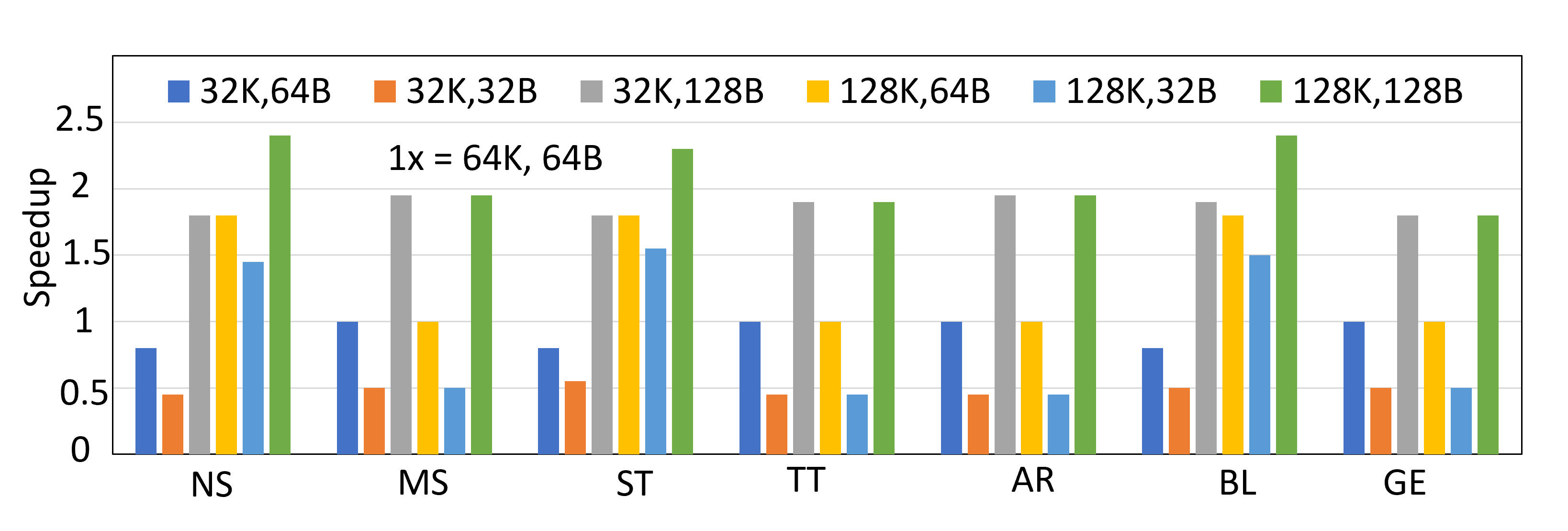}
\vspace{-2em}
\caption{\textbf{ Speedup normalized to 64KB L1 cache, read of 64B. Label X,Y implies L1 cache of size X, read of Y }}
\label{Cache_size}
\vspace{-1em}
\end{figure}
\subsection{SPARK's performance with varying L1/L2/L3 sizes}
Fig.\ref{Cache_size} shows SPARK's performance variation with L1 size, highlighting the interaction between L1 cache size and cache throughput (determined by read/compute size).

\par \underline{Cache size}: For workloads that do not fit in the L1 cache (such as NS, ST, and BL), reducing the size of the L1 cache further leads to a decrease in performance. This is due to the increased number of cache misses and the need to fetch data from slower levels of memory. However, prefetching mitigates most of this performance loss, limiting the overall performance drop to just 0.2x. On the other hand, increasing the L1 cache size for these workloads provides a performance boost, achieving a speedup of up to 1.5x by reducing cache misses and improving memory access times. For workloads that fit entirely within the L1 cache, performance remains unaffected by changes in cache size, as all necessary data is already available in the faster L1 cache.
\par \underline{Read/compute size}: L1 cache read/compute size is crucial as it directly affects SPARK's near-memory logic utilization. Halving the throughput typically reduces speedup by half across most workloads. Conversely, doubling throughput, size enables speedups of 2.4x for workloads that don’t fit in L1.
\par For L2 cache, read/compute size is irrelevant since we don’t perform near-L2 compute (reserved for follow-up work). Reducing L2 size from 4MB to 2MB halves performance, but prefetching can recover up to 0.9x. L3 cache shows no sensitivity in MIPLIB benchmarks, as they fit within L1, L2.
\section{Discussion}
\subsection{SPARK's importance}
Firstly, SPARK accelerates ILP/LP with minimal area overhead by tightly integrating CPUs with L1 cache, exceeding decision thresholds in several benchmarks (TT, AR, BL, GE). It delivers significant speedups over CPU/GPU and lower energy compared to existing methods. Even when the threshold isn't met for some benchmarks, SPARK consumes less energy than CPU/GPU execution, enabling energy-efficient ILP execution. We have developed near-L1, L2 compute, which improves results in benchmarks that doesn't meet threshold, left for future work. Thus, SPARK is a high-performance, energy-efficient architecture and a foundation for other accelerators.
\par Secondly, there are 2 broadly used architectures for accelerating workloads, chosen based on the workload: 
\\ i) \underline{Dedicated accelerator} - Located farther from the CPU, these are used for specialized workloads, where the cost of moving data is offset by the accelerator’s high throughput.
\\ ii) \underline{GPU} - Depending on the SoC design and programming model, GPU may take precedence of execution over dedicated accelerators but offers similar tradeoffs as that of dedicated accelerators in terms of throughput/data movement.
\par In SPARK, we demonstrate that workloads like ILP require real-time processing, making near L1-cache compute necessary for energy-efficient, high-performance design. Thus, we propose an architecture tightly integrated with CPU's L1, reconfiguring CPU's L1 for compute.
\subsection{SPARK's generality to different ILP algorithms}
\par SPARK is adaptable to various ILP algorithms, allowing them to be mapped onto it without hardware modifications. In ILP, constraints are typically expressed as $C*X \leq D $, and different algorithms use this to solve problems efficiently.
\par For a \underline{small number of constraints} where direct methods like Cramer's rule may be preferred, the task is to solve \(n\)-equations with \(n\)-variables and check if other equations satisfy the solution. For example, consider 3 constraints of the form \( C_{ij} * X_j \leq D_i \). Coefficients are stored in memory, with RBL mapped to \( C_{ij} \) to compute the bit-wise dot-product. Using SA and AR, the final dot-product between \( C_{ii} \) and \( C_{ij} \) is written into queues. Subtraction is then performed by reading from the queues, followed by division to get the final result.
\par For a \underline{larger problem}, hypothetically if Gauss-Seidel method \cite{Gauss-seidel} is preferred over Jacobi, the lower and upper triangular matrices are stored in memory. \( X \) is multiplied by the upper triangular matrix as in Jacobi, and \( D \) is subtracted similarly. The divider computes the determinant, and multiplication with the lower triangular matrix follows same approach as Jacobi.

\subsection{Algorithmic insight and regularizing divider for low area}
\par \underline{Background}: 
Jacobi's iterative method and B\&B seek local optima using the L1 norm but often get stuck in local minima, mitigated by annealing or regularization. Algorithmically, division is the final step in each iteration, followed by regularization, requiring extra hardware. Hardware-wise, dividers are area-intensive and not ideal for near-memory compute.

\par \underline{Idea}: We propose using a "regularizing-divider" for regularization by employing less-accurate division through approximate dividers, as in \cite{CADE}\cite{Dynamic_div}. This replaces costly division with subtracting the first m-bits of mantissa values, with m adjustable based on error. If the error exceeds 1\%, a 64B lookup table (shared across memory banks) provides a correction value to refine the subtraction for the final output.
\par \underline{Results}: Subtraction approximation reduces gate depth, achieving 0.5ns latency, 0.15pJ energy in FreePDK 45nm technology, enabling energy-efficient single-cycle division. This results in an average error of 0.2\% on MIPLIB benchmarks, aiding regularization of updated X values. The lookup table occupies 0.02mm\textsuperscript{2} per core, with a subtractor area of 0.04mm\textsuperscript{2}. 

\subsection{Memory design}
This architecture \cite{SPARK} does not make any assumptions about the underlying memory design and the compute architecture can be implemented using any form of emerging non-volatile memory including Ferroelectric Field effect transistors (FeFET) \cite{fefet}, embedded DRAM \cite{Shanshan_Ising}\cite{SACHI_arxiv}\cite{DRAM_PIM}\cite{ABI}\cite{Ising_review}\cite{UT_Thesis}, non-volatile SRAM \cite{8T_SRAM_1}, resistive random access memory \cite{NVM_Raman}\cite{RRAM_cache}. 

\subsection{Operating temperature}
Furthermore, this work does not make any assumptions about the operational temperature, and can operate in cryogenic temperatures as well exploiting memory array designs like the flop-arrays mentioned in \cite{UTBB_SOI}, JJFET mentioned in \cite{JJFET}\cite{Cryo_arxiv}
\section{CONCLUSION}
We propose Spark, a near-memory sparsity-aware accelerator that reconfigures L1 cache in CPUs for high-throughput compute, removing redundant computations with a  sparsity-aware approach, reuse-aware approach for control-flow intensive operations that helps reduce the near-memory logic area overhead. 
Spark achieves speedup of 15x/20x and energy benefits of 152x/740x, over AMD's Zen3 CPU/Nvidia's Tesla v100 GPU for real-life sparse MIPLIB 2017 applications. In dense ILPs, Spark achieves 6-10x/60-210x performance/energy improvement over CPU and GPU. Spark achieves 7-17x/103-250x performance/energy improvement over CPU and GPU in sparse LP, and 5-7x/150-180x performance/energy improvement over CPU/GPU in dense LP.  {\color{red} }    

\bibliographystyle{IEEEtranS}
\bibliography{refs}

@misc{ABI,
      title={ABI: A tightly integrated, unified, sparsity-aware, reconfigurable, compute near-register file/cache GPU architecture with light-weight softmax for deep learning, linear algebra, and Ising compute}, 
      author={Siddhartha Raman Sundara Raman and Jaydeep P. Kulkarni},
      year={2026},
      eprint={2602.14262},
      archivePrefix={arXiv},
      primaryClass={cs.AR},
      url={https://arxiv.org/abs/2602.14262}, 
}

@article{DRAM_PIM,
  title={A comparative study on power delivery aspects of compute-in/near-memory approaches using DRAM},
  author={Raman, Siddhartha Raman Sundara and Ma, Siyuan and John, Lizy Kurian},
  journal={arXiv preprint arXiv:2604.04773},
  year={2026}
}

@INPROCEEDINGS{SPARK,
  author={Raman, Siddhartha Raman Sundara and John, Lizy and Kulkarni, Jaydeep P.},
  booktitle={2025 IEEE International Symposium on High Performance Computer Architecture (HPCA)}, 
  title={SPARK: Sparsity Aware, Low Area, Energy-Efficient, Near-memory Architecture for Accelerating Linear Programming Problems}, 
  year={2025},
  volume={},
  number={},
  pages={99-112},
  doi={10.1109/HPCA61900.2025.00019}}

@ARTICLE{Ising_review,
  author={Kulkarni, Jaydeep P. and Sundara Raman, Siddhartha Raman and Xie, Shanshan and Lo, Chieh-Pu},
  journal={Computer}, 
  title={Unconventional Computing Using Ising Accelerators}, 
  year={2025},
  volume={58},
  number={6},
  pages={83-86},
  doi={10.1109/MC.2025.3544798}}

@ARTICLE{JJFET,
  author={Raman, Siddhartha Raman Sundara and Wen, Feng and Pillarisetty, Ravi and De, Vivek and Kulkarni, Jaydeep P.},
  journal={IEEE Transactions on Applied Superconductivity}, 
  title={High Noise Margin, Digital Logic Design Using Josephson Junction Field-Effect Transistors for Cryogenic Computing}, 
  year={2021},
  volume={31},
  number={5},
  pages={1-5},
  doi={10.1109/TASC.2021.3054347}}

@article{RRAM_cache,
    author = {Boppidi, Pavan Kumar Reddy and Raman, S. Siddhartha and Renuka, H. and Kundu, Souvik},
    title = "{Pt/Cu:ZnO/Nb:STO memristive dual port for cache memory applications}",
    journal = {AIP Conference Proceedings},
    volume = {2265},
    number = {1},
    pages = {030212},
    year = {2020},
    month = {11},
    issn = {0094-243X},
    doi = {10.1063/5.0016597},
    url = {https://doi.org/10.1063/5.0016597},
    eprint = {https://pubs.aip.org/aip/acp/article-pdf/doi/10.1063/5.0016597/14105127/030212\_1\_online.pdf},
}

@ARTICLE{UTBB_SOI,
  author={Nibhanupudi, S. S. Teja and Sundara Raman, Siddhartha Raman and Cassé, Mikaël and Hutin, Louis and Kulkarni, Jaydeep P.},
  journal={IEEE Journal on Exploratory Solid-State Computational Devices and Circuits}, 
  title={Ultra-Low-Voltage UTBB-SOI-Based, Pseudo-Static Storage Circuits for Cryogenic CMOS Applications}, 
  year={2021},
  volume={7},
  number={2},
  pages={201-208},
  doi={10.1109/JXCDC.2021.3130839}}

@incollection{NVM_Raman,
author = {Siddhartha Raman Sundara Raman},
title = {A Review on Non-Volatile and Volatile Emerging Memory Technologies},
booktitle = {Computer Memory and Data Storage},
publisher = {IntechOpen},
address = {Rijeka},
year = {2024},
editor = {Azam Seyedi},
chapter = {3},
doi = {10.5772/intechopen.110617},
url = {https://doi.org/10.5772/intechopen.110617}
}

@ARTICLE{fefet,
  author={Raman, Siddhartha Raman Sundara and Nibhanupudi, S. S. Teja and Saha, Atanu K. and Gupta, Sumeet and Kulkarni, Jaydeep P.},
  journal={IEEE Transactions on Electron Devices}, 
  title={Threshold Selector and Capacitive Coupled Assist Techniques for Write Voltage Reduction in Metal–Ferroelectric–Metal Field-Effect Transistor}, 
  year={2021},
  volume={68},
  number={12},
  pages={6132-6138},
  doi={10.1109/TED.2021.3121348}}

@article{TACO,
author = {Sundara Raman, Siddhartha Raman and John, Lizy and Kulkarni, Jaydeep P.},
title = {NEM-GNN: DAC/ADC-less, Scalable, Reconfigurable, Graph and Sparsity-Aware Near-Memory Accelerator for Graph Neural Networks},
year = {2024},
issue_date = {June 2024},
publisher = {Association for Computing Machinery},
address = {New York, NY, USA},
volume = {21},
number = {2},
issn = {1544-3566},
url = {https://doi.org/10.1145/3652607},
doi = {10.1145/3652607},
journal = {ACM Trans. Archit. Code Optim.},
month = may,
articleno = {39},
numpages = {26}
}

@inproceedings{CADE,
  title={Cade: Configurable approximate divider for energy efficiency},
  author={Imani, Mohsen and Garcia, Ricardo and Huang, Andrew and Rosing, Tajana},
  booktitle={2019 Design, Automation \& Test in Europe Conference \& Exhibition (DATE)},
  pages={586--589},
  year={2019},
  organization={IEEE}
}

@inproceedings{Dynamic_div,
  title={A low-power dynamic divider for approximate applications},
  author={Hashemi, Soheil and Bahar, R Iris and Reda, Sherief},
  booktitle={Proceedings of the 53rd Annual Design Automation Conference},
  pages={1--6},
  year={2016}
}

@article{Tensi,
  title={Tensilica CPU bends to designers’ will},
  author={Turley, Jim},
  journal={Microprocessor Report},
  volume={13},
  number={3},
  pages={12},
  year={1999}
}

@inproceedings{Compute_cache,
  title={Compute caches},
  author={Aga, Shaizeen and Jeloka, Supreet and Subramaniyan, Arun and Narayanasamy, Satish and Blaauw, David and Das, Reetuparna},
  booktitle={2017 IEEE International Symposium on High Performance Computer Architecture (HPCA)},
  pages={481--492},
  year={2017},
  organization={IEEE}
}

@phdthesis{UT_Thesis,
  author       = {Siddhartha Raman Sundara Raman},
  title        = {Compute in eDRAM using indium gallium zinc oxide transistors},
  school       = {The University of Texas at Austin},
  year         = {2026},
  type         = {Ph.D. dissertation},
  url          = {https://repositories.lib.utexas.edu/items/4dbc7f92-c062-4cb8-b07b-ed29761b9704},
  note         = {Available: \url{https://repositories.lib.utexas.edu/items/4dbc7f92-c062-4cb8-b07b-ed29761b9704}}
}

@misc{NEM_GNN_arxiv,
      title={A complete discussion on fully reconfigurable, digital, scalable, graph and sparsity-aware near-memory accelerator for graph neural networks}, 
      author={Siddhartha Raman Sundara Raman and Lizy John and Jaydeep P. Kulkarni},
      note={arXiv preprint arXiv:2605.19405},
      url={https://arxiv.org/abs/2605.19405}, 
}

@misc{Cryo_arxiv,
      title={Emerging memory technologies at room/cryogenic temperature}, 
      author={Siddhartha Raman Sundara Raman},
      year={2026},
      note={arXiv preprint arXiv:2605.21912},
      url={https://arxiv.org/abs/2605.21912}, 
}

@INPROCEEDINGS{SACHI,
  author={Sundara Raman, Siddhartha Raman and John, Lizy K. and Kulkarni, Jaydeep P.},
  booktitle={2024 IEEE International Symposium on High-Performance Computer Architecture (HPCA)}, 
  title={SACHI: A Stationarity-Aware, All-Digital, Near-Memory, Ising Architecture}, 
  year={2024},
  volume={},
  number={},
  pages={719-731},
  doi={10.1109/HPCA57654.2024.00061}}

@misc{CloudTPU,
  title = {[Online] Introduction to Cloud TPU},
  howpublished = {\url{https://cloud.google.com/tpu/docs/intro-to-tpu}},
  
}

@inproceedings{8T_L1_2,
  title={Dual-V CC 8T-bitcell SRAM array in 22nm tri-gate CMOS for energy-efficient operation across wide dynamic voltage range},
  author={Kulkarni, Jaydeep and Khellah, Muhammad and Tschanz, Jim and Geuskens, Bibiche and Jain, Rinkle and Kim, Stephen and De, Vivek},
  booktitle={2013 Symposium on VLSI Technology},
  pages={C126--C127},
  year={2013},
  organization={IEEE}
}

@article{8T_L1,
  title={Wide-Range Many-Core SoC Design in Scaled CMOS: Challenges and Opportunities},
  author={Vangal, Sriram and Paul, Somnath and Hsu, Steven and Agarwal, Amit and Kumar, Saurabh and Krishnamurthy, Ram and Krishnamurthy, Harish and Tschanz, James and De, Vivek and Kim, Chris H},
  journal={IEEE Transactions on Very Large Scale Integration (VLSI) Systems},
  volume={29},
  number={5},
  pages={843--856},
  year={2021},
  publisher={IEEE}
}

@article{RNS,
  title={Design of a Residue Number System Based Linear System Solver in Hardware},
  author={Bu{\v{c}}ek, Ji{\v{r}}{\'\i} and Kubal{\'\i}k, Pavel and L{\'o}rencz, R{\'o}bert and Zahradnick{\`y}, Tom{\'a}{\v{s}}},
  journal={Journal of Signal Processing Systems},
  volume={87},
  pages={343--356},
  year={2017},
  publisher={Springer}
}

@article{Survey,
  title={A Survey of Design and Optimization for Systolic Array Based DNN Accelerators},
  author={Xu, Rui and Ma, Sheng and Guo, Yang and Li, Dongsheng},
  journal={ACM Computing Surveys},
  year={2023},
  publisher={ACM New York, NY}
}

@article{Ising,
  title={Ising formulations of many NP problems},
  author={Lucas, Andrew},
  journal={Frontiers in physics},
  volume={2},
  pages={5},
  year={2014},
  publisher={Frontiers}
}

@article{EIE,
  title={EIE: Efficient inference engine on compressed deep neural network},
  author={Han, Song and Liu, Xingyu and Mao, Huizi and Pu, Jing and Pedram, Ardavan and Horowitz, Mark A and Dally, William J},
  journal={ACM SIGARCH Computer Architecture News},
  volume={44},
  number={3},
  pages={243--254},
  year={2016},
  publisher={ACM New York, NY, USA}
}

@inproceedings{Sparse_TPU,
  title={Sparse-TPU: Adapting systolic arrays for sparse matrices},
  author={He, Xin and Pal, Subhankar and Amarnath, Aporva and Feng, Siying and Park, Dong-Hyeon and Rovinski, Austin and Ye, Haojie and Chen, Yuhan and Dreslinski, Ronald and Mudge, Trevor},
  booktitle={Proceedings of the 34th ACM international conference on supercomputing},
  pages={1--12},
  year={2020}
}

@misc{CCF,
  title={Ccf: A cgra compilation framework},
  author={Dave, Shail and Shrivastava, Aviral},
  year={2018}
}

@inproceedings{Transmuter,
  title={Transmuter: Bridging the efficiency gap using memory and dataflow reconfiguration},
  author={Pal, Subhankar and Feng, Siying and Park, Dong-hyeon and Kim, Sung and Amarnath, Aporva and Yang, Chi-Sheng and He, Xin and Beaumont, Jonathan and May, Kyle and Xiong, Yan and others},
  booktitle={Proceedings of the ACM International Conference on Parallel Architectures and Compilation Techniques},
  pages={175--190},
  year={2020}
}

@inproceedings{Bison-e,
  title={Bison-e: A lightweight and high-performance accelerator for narrow integer linear algebra computing on the edge},
  author={Reggiani, Enrico and Lazo, Crist{\'o}bal Ram{\'\i}rez and Bagu{\'e}, Roger Figueras and Cristal, Adri{\'a}n and Olivieri, Mauro and Unsal, Osman Sabri},
  booktitle={Proceedings of the 27th ACM International Conference on Architectural Support for Programming Languages and Operating Systems},
  pages={56--69},
  year={2022}
}

@article{CPLEX,
  title={V12. 1: User’s Manual for CPLEX},
  author={Cplex, IBM ILOG},
  journal={International Business Machines Corporation},
  volume={46},
  number={53},
  pages={157},
  year={2009}
}

@inproceedings{tpu-v4,
author = {Jouppi, Norm and Kurian, George and Li, Sheng and Ma, Peter and Nagarajan, Rahul and Nai, Lifeng and Patil, Nishant and Subramanian, Suvinay and Swing, Andy and Towles, Brian and Young, Clifford and Zhou, Xiang and Zhou, Zongwei and Patterson, David A},
title = {TPU v4: An Optically Reconfigurable Supercomputer for Machine Learning with Hardware Support for Embeddings},
year = {2023},
isbn = {9798400700958},
publisher = {Association for Computing Machinery},
address = {New York, NY, USA},
url = {https://doi.org/10.1145/3579371.3589350},
doi = {10.1145/3579371.3589350},
booktitle = {Proceedings of the 50th Annual International Symposium on Computer Architecture},
articleno = {82},
numpages = {14},
location = {Orlando, FL, USA},
series = {ISCA '23}
}

@article{tpu-v1,
author = {Jouppi, Norman P. and Young, Cliff and Patil, Nishant and Patterson, David and Agrawal, Gaurav and Bajwa, Raminder and Bates, Sarah and Bhatia, Suresh and Boden, Nan and Borchers, Al and Boyle, Rick and Cantin, Pierre-luc and Chao, Clifford and Clark, Chris and Coriell, Jeremy and Daley, Mike and Dau, Matt and Dean, Jeffrey and Gelb, Ben and Ghaemmaghami, Tara Vazir and Gottipati, Rajendra and Gulland, William and Hagmann, Robert and Ho, C. Richard and Hogberg, Doug and Hu, John and Hundt, Robert and Hurt, Dan and Ibarz, Julian and Jaffey, Aaron and Jaworski, Alek and Kaplan, Alexander and Khaitan, Harshit and Killebrew, Daniel and Koch, Andy and Kumar, Naveen and Lacy, Steve and Laudon, James and Law, James and Le, Diemthu and Leary, Chris and Liu, Zhuyuan and Lucke, Kyle and Lundin, Alan and MacKean, Gordon and Maggiore, Adriana and Mahony, Maire and Miller, Kieran and Nagarajan, Rahul and Narayanaswami, Ravi and Ni, Ray and Nix, Kathy and Norrie, Thomas and Omernick, Mark and Penukonda, Narayana and Phelps, Andy and Ross, Jonathan and Ross, Matt and Salek, Amir and Samadiani, Emad and Severn, Chris and Sizikov, Gregory and Snelham, Matthew and Souter, Jed and Steinberg, Dan and Swing, Andy and Tan, Mercedes and Thorson, Gregory and Tian, Bo and Toma, Horia and Tuttle, Erick and Vasudevan, Vijay and Walter, Richard and Wang, Walter and Wilcox, Eric and Yoon, Doe Hyun},
title = {In-Datacenter Performance Analysis of a Tensor Processing Unit},
year = {2017},
issue_date = {May 2017},
publisher = {Association for Computing Machinery},
address = {New York, NY, USA},
volume = {45},
number = {2},
issn = {0163-5964},
url = {https://doi.org/10.1145/3140659.3080246},
doi = {10.1145/3140659.3080246},
journal = {SIGARCH Comput. Archit. News},
month = {jun},
pages = {1–12},
numpages = {12}
}

@article{guorobi-blog,
  title={Guorobi Blog},
  author={Greg Gockner},
  journal={https://support.gurobi.com/hc/en-us/articles/360012237852-Does-Gurobi-support-GPUs-},
  volume={},
  number={},
  pages={},
  year={2023}
}

@inproceedings{GoSpa,
  title={Gospa: An energy-efficient high-performance globally optimized sparse convolutional neural network accelerator},
  author={Deng, Chunhua and Sui, Yang and Liao, Siyu and Qian, Xuehai and Yuan, Bo},
  booktitle={2021 ACM/IEEE 48th Annual International Symposium on Computer Architecture (ISCA)},
  pages={1110--1123},
  year={2021},
  organization={IEEE}
}

@inproceedings{Extensor,
  title={Extensor: An accelerator for sparse tensor algebra},
  author={Hegde, Kartik and Asghari-Moghaddam, Hadi and Pellauer, Michael and Crago, Neal and Jaleel, Aamer and Solomonik, Edgar and Emer, Joel and Fletcher, Christopher W},
  booktitle={Proceedings of the 52nd Annual IEEE/ACM International Symposium on Microarchitecture},
  pages={319--333},
  year={2019}
}

@inproceedings{SparTen,
author = {Gondimalla, Ashish and Chesnut, Noah and Thottethodi, Mithuna and Vijaykumar, T. N.},
title = {SparTen: A Sparse Tensor Accelerator for Convolutional Neural Networks},
year = {2019},
isbn = {9781450369381},
publisher = {Association for Computing Machinery},
address = {New York, NY, USA},
url = {https://doi.org/10.1145/3352460.3358291},
doi = {10.1145/3352460.3358291},
booktitle = {Proceedings of the 52nd Annual IEEE/ACM International Symposium on Microarchitecture},
pages = {151–165},
numpages = {15},
location = {Columbus, OH, USA},
series = {MICRO '52}
}

@INPROCEEDINGS{Horowitz,
  author={Horowitz, Mark},
  booktitle={2014 IEEE International Solid-State Circuits Conference Digest of Technical Papers (ISSCC)}, 
  title={1.1 Computing's energy problem (and what we can do about it)}, 
  year={2014},
  volume={},
  number={},
  pages={10-14},
  doi={10.1109/ISSCC.2014.6757323}}

@article{Jacobi,
  title={The Hamilton-Jacobi method and Hamiltonian maps},
  author={Abdullaev, SS},
  journal={Journal of Physics A: Mathematical and General},
  volume={35},
  number={12},
  pages={2811},
  year={2002},
  publisher={IOP Publishing}
}

@article{Gauss-seidel,
  title={Adaptive Gauss-Seidel method for linear systems},
  author={Usui, Masataka and Niki, Hiroshi and Kohno, Toshiyuki},
  journal={International Journal of Computer Mathematics},
  volume={51},
  number={1-2},
  pages={119--125},
  year={1994},
  publisher={Taylor \& Francis}
}

@article{Nelder-mead,
  title={Implementing the Nelder-Mead simplex algorithm with adaptive parameters},
  author={Gao, Fuchang and Han, Lixing},
  journal={Computational Optimization and Applications},
  volume={51},
  number={1},
  pages={259--277},
  year={2012},
  publisher={Springer}
}

@article{Simplex,
  title={How good is the simplex algorithm},
  author={Klee, Victor and Minty, George J},
  journal={Inequalities},
  volume={3},
  number={3},
  pages={159--175},
  year={1972},
  publisher={New York}
}

@ARTICLE{6T_SRAM,  author={Nibhanupudi, S. S. Teja and Raman, Siddhartha Raman Sundara and Kulkarni, Jaydeep P.},  journal={IEEE Transactions on Electron Devices},   title={Phase Transition Material-Assisted Low-Power SRAM Design},   year={2021},  volume={68},  number={5},  pages={2281-2288},  doi={10.1109/TED.2021.3067849}}

@incollection{Symmetric_ILP_1,
  title={Symmetry in mathematical programming},
  author={Liberti, Leo},
  booktitle={Mixed Integer Nonlinear Programming},
  pages={263--283},
  year={2012},
  publisher={Springer}
}

@article{Symmetric_ILP,
  title={Exploiting orbits in symmetric ILP},
  author={Margot, Fran{\c{c}}ois},
  journal={Mathematical Programming},
  volume={98},
  number={1},
  pages={3--21},
  year={2003},
  publisher={Springer}
}

@book{LP_book,
  title={Linear programming},
  author={Chvatal, Vasek and Chvatal, Vaclav and others},
  year={1983},
  publisher={Macmillan}
}

@article{LP,
  title={Linear programming},
  author={Dantzig, George B},
  journal={Operations research},
  volume={50},
  number={1},
  pages={42--47},
  year={2002},
  publisher={INFORMS}
}

@article{Jacobi_iterative,
  title={A Jacobi--Davidson iteration method for linear eigenvalue problems},
  author={Sleijpen, Gerard LG and Van der Vorst, Henk A},
  journal={SIAM review},
  volume={42},
  number={2},
  pages={267--293},
  year={2000},
  publisher={SIAM}
}

@INPROCEEDINGS{eDRAM_1,  author={Raman, Siddhartha Raman Sundara and Xie, Shanshan and P.Kulkarni, Jaydeep},  booktitle={2021 IEEE International Symposium on Circuits and Systems (ISCAS)},   title={Compute-in-eDRAM with Backend Integrated Indium Gallium Zinc Oxide Transistors},   year={2021},  volume={},  number={},  pages={1-5},  doi={10.1109/ISCAS51556.2021.9401798}}

@ARTICLE{8T_SRAM_1,  author={Sundara Raman, Siddhartha Raman and Nibhanupudi, S. S. Teja and Kulkarni, Jaydeep P.},  journal={IEEE Journal on Emerging and Selected Topics in Circuits and Systems},   title={Enabling In-Memory Computations in Non-Volatile SRAM Designs},   year={2022},  volume={12},  number={2},  pages={557-568},  doi={10.1109/JETCAS.2022.3174148}}

@inproceedings{Branch_div,
  title={Reducing branch divergence in GPU programs},
  author={Han, Tianyi David and Abdelrahman, Tarek S},
  booktitle={Proceedings of the fourth workshop on general purpose processing on graphics processing units},
  pages={1--8},
  year={2011}
}

@inproceedings{ILP_solver,
  title={Generic ILP versus specialized 0-1 ILP: An update},
  author={Aloul, Fadi A and Ramani, Arathi and Markov, Igor L and Sakallah, Karem A},
  booktitle={Proceedings of the 2002 IEEE/ACM international conference on Computer-aided design},
  pages={450--457},
  year={2002}
}

@INPROCEEDINGS{FPGA_ILP,  author={Bayliss, Samuel and Bouganis, Christos-s. and Constantinides, George A. and Luk, Wayne},  booktitle={2006 IEEE International Conference on Field Programmable Technology},   title={An FPGA implementation of the simplex algorithm},   year={2006},  volume={},  number={},  pages={49-56},  doi={10.1109/FPT.2006.270294}}

@INPROCEEDINGS{SLE_SOC,  author={BuČek, Jiří and Kubalík, Pavel and Lórencz, Róbert and Zahradnický, Tomáš},  booktitle={2014 International Symposium on System-on-Chip (SoC)},   title={System on chip design of a linear system solver},   year={2014},  volume={},  number={},  pages={1-6},  doi={10.1109/ISSOC.2014.6972445}}

@incollection{B&B_ILP,
  title={Heuristic analysis, linear programming and branch and bound},
  author={Wolsey, Laurence A},
  booktitle={Combinatorial Optimization II},
  pages={121--134},
  year={1980},
  publisher={Springer}
}

@incollection{Acad_ILP,
  title={Progress in academic computational integer programming},
  author={Koch, Thorsten and Martin, Alexander and Pfetsch, Marc E},
  booktitle={Facets of Combinatorial Optimization},
  pages={483--506},
  year={2013},
  publisher={Springer}
}

@article{Data_Driven_library,
  title={MIPLIB 2017: data-driven compilation of the 6th mixed-integer programming library},
  author={Gleixner, Ambros and Hendel, Gregor and Gamrath, Gerald and Achterberg, Tobias and Bastubbe, Michael and Berthold, Timo and Christophel, Philipp and Jarck, Kati and Koch, Thorsten and Linderoth, Jeff and others},
  journal={Mathematical Programming Computation},
  volume={13},
  number={3},
  pages={443--490},
  year={2021},
  publisher={Springer}
}

@article{MIP,
  title={A computational study of primal heuristics inside an MI (NL) P solver},
  author={Berthold, Timo},
  journal={Journal of Global Optimization},
  volume={70},
  number={1},
  pages={189--206},
  year={2018},
  publisher={Springer}
}

@article{MIPLIB_2017,
  author                   = {Gleixner, Ambros and Hendel, Gregor and Gamrath, Gerald and Achterberg, Tobias and Bastubbe, Michael and Berthold, Timo and Christophel, Philipp M. and Jarck, Kati and Koch, Thorsten and Linderoth, Jeff and L\"ubbecke, Marco and Mittelmann, Hans D. and Ozyurt, Derya and Ralphs, Ted K. and Salvagnin, Domenico and Shinano, Yuji},
  title                    = {{MIPLIB 2017: Data-Driven Compilation of the 6th Mixed-Integer Programming Library}},
  journal                  = {Mathematical Programming Computation},
  year                     = {2021},
  doi                      = {10.1007/s12532-020-00194-3},
  url                      = {https://doi.org/10.1007/s12532-020-00194-3}
}

@article{Solvers_2020,
  title={Progress in mathematical programming solvers from 2001 to 2020},
  author={Koch, Thorsten and Berthold, Timo and Pedersen, Jaap and Vanaret, Charlie},
  journal={EURO Journal on Computational Optimization},
  pages={100031},
  year={2022},
  publisher={Elsevier}
}

@article{Simplex_1,
  title={Parallelizing the dual revised simplex method},
  author={Huangfu, Qi and Hall, JA Julian},
  journal={Mathematical Programming Computation},
  volume={10},
  number={1},
  pages={119--142},
  year={2018},
  publisher={Springer}
}

@inproceedings{wave-pim,
  title={Wave-pim: Accelerating wave simulation using processing-in-memory},
  author={Hanindhito, Bagus and Li, Ruihao and Gourounas, Dimitrios and Fathi, Arash and Govil, Karan and Trenev, Dimitar and Gerstlauer, Andreas and John, Lizy},
  booktitle={Proceedings of the 50th International Conference on Parallel Processing},
  pages={1--11},
  year={2021}
}

@inproceedings{comefa,
  title={CoMeFa: Compute-in-Memory Blocks for FPGAs},
  author={Arora, Aman and Anand, Tanmay and Borda, Aatman and Sehgal, Rishabh and Hanindhito, Bagus and Kulkarni, Jaydeep and John, Lizy K},
  booktitle={2022 IEEE 30th Annual International Symposium on Field-Programmable Custom Computing Machines (FCCM)},
  pages={1--9},
  year={2022},
  organization={IEEE}
}

@inproceedings{neuralcache,
  title={Neural cache: Bit-serial in-cache acceleration of deep neural networks},
  author={Eckert, Charles and Wang, Xiaowei and Wang, Jingcheng and Subramaniyan, Arun and Iyer, Ravi and Sylvester, Dennis and Blaaauw, David and Das, Reetuparna},
  booktitle={2018 ACM/IEEE 45Th annual international symposium on computer architecture (ISCA)},
  pages={383--396},
  year={2018},
  organization={IEEE}
}

@article{B_B_random,
  title={Improving branch-and-cut performance by random sampling},
  author={Fischetti, Matteo and Lodi, Andrea and Monaci, Michele and Salvagnin, Domenico and Tramontani, Andrea},
  journal={Mathematical Programming Computation},
  volume={8},
  number={1},
  pages={113--132},
  year={2016},
  publisher={Springer}
}

@article{CPUs_ILP,
  title={Could we use a million cores to solve an integer program?},
  author={Koch, Thorsten and Ralphs, Ted and Shinano, Yuji},
  journal={Mathematical Methods of Operations Research},
  volume={76},
  number={1},
  pages={67--93},
  year={2012},
  publisher={Springer}
}

@ARTICLE{Fast_branch_bound,  author={Somol, P. and Pudil, P. and Kittler, J.},  journal={IEEE Transactions on Pattern Analysis and Machine Intelligence},   title={Fast branch \& bound algorithms for optimal feature selection},   year={2004},  volume={26},  number={7},  pages={900-912},  doi={10.1109/TPAMI.2004.28}}

@article{Thread_divergence_GPU,
  title={Reducing thread divergence in a GPU-accelerated branch-and-bound algorithm},
  author={Chakroun, Imen and Mezmaz, Mohand and Melab, Nouredine and Bendjoudi, Ahcene},
  journal={Concurrency and Computation: Practice and Experience},
  volume={25},
  number={8},
  pages={1121--1136},
  year={2013},
  publisher={Wiley Online Library}
}

@misc{SACHI_arxiv,
      title={A detailed algorithmic study on a reuse-aware, near memory, all-digital Ising machine}, 
      author={Siddhartha Raman Sundara Raman and Lizy K. John and Jaydeep P. Kulkarni},
      year={2026},
      eprint={2605.12959},
      archivePrefix={arXiv},
      primaryClass={cs.AR},
      url={https://arxiv.org/abs/2605.12959}, 
}

@article{8T_SRAM,
  title={An 8T-SRAM for variability tolerance and low-voltage operation in high-performance caches},
  author={Chang, Leland and Montoye, Robert K and Nakamura, Yutaka and Batson, Kevin A and Eickemeyer, Richard J and Dennard, Robert H and Haensch, Wilfried and Jamsek, Damir},
  journal={IEEE Journal of Solid-State Circuits},
  volume={43},
  number={4},
  pages={956--963},
  year={2008},
  publisher={IEEE}
}

@inproceedings{45nmPDK,
  title={FreePDK: An open-source variation-aware design kit},
  author={Stine, James E and Castellanos, Ivan and Wood, Michael and Henson, Jeff and Love, Fred and Davis, W Rhett and Franzon, Paul D and Bucher, Michael and Basavarajaiah, Sunil and Oh, Julie and others},
  booktitle={2007 IEEE international conference on Microelectronic Systems Education (MSE'07)},
  pages={173--174},
  year={2007},
  organization={IEEE}
}

@inproceedings{SARA,
  title={Sara: Scaling a reconfigurable dataflow accelerator},
  author={Zhang, Yaqi and Zhang, Nathan and Zhao, Tian and Vilim, Matt and Shahbaz, Muhammad and Olukotun, Kunle},
  booktitle={2021 ACM/IEEE 48th Annual International Symposium on Computer Architecture (ISCA)},
  pages={1041--1054},
  year={2021},
  organization={IEEE}
}

@inproceedings{Cosa,
  title={Cosa: Scheduling by constrained optimization for spatial accelerators},
  author={Huang, Qijing and Kang, Minwoo and Dinh, Grace and Norell, Thomas and Kalaiah, Aravind and Demmel, James and Wawrzynek, John and Shao, Yakun Sophia},
  booktitle={2021 ACM/IEEE 48th Annual International Symposium on Computer Architecture (ISCA)},
  pages={554--566},
  year={2021},
  organization={IEEE}
}

@misc{gurobi,
  author = {{Gurobi Optimization, LLC}},
  title = {{Gurobi Optimizer Reference Manual}},
  year = 2022,
  url = "https://www.gurobi.com"
}

@ARTICLE{IGZO_CIM,  author={Sundara Raman, Siddhartha Raman and Xie, Shanshan and Kulkarni, Jaydeep P.},  journal={IEEE Journal on Exploratory Solid-State Computational Devices and Circuits},   title={IGZO CIM: Enabling In-Memory Computations Using Multilevel Capacitorless Indium–Gallium–Zinc–Oxide-Based Embedded DRAM Technology},   year={2022},  volume={8},  number={1},  pages={35-43},  doi={10.1109/JXCDC.2022.3188366}}

@ARTICLE{Shanshan_Ising,  author={Xie, Shanshan and Raman, Siddhartha Raman Sundara and Ni, Can and Wang, Meizhi and Yang, Mengtian and Kulkarni, Jaydeep P.},  journal={IEEE Journal of Solid-State Circuits},   title={Ising-CIM: A Reconfigurable and Scalable Compute Within Memory Analog Ising Accelerator for Solving Combinatorial Optimization Problems},   year={2022},  volume={},  number={},  pages={1-13},  doi={10.1109/JSSC.2022.3176610}}

@INPROCEEDINGS{CIM-Spin,  author={Su, Yuqi and Kim, Hyunjoon and Kim, Bongjin},  booktitle={2020 IEEE International Solid- State Circuits Conference - (ISSCC)},   title={31.2 CIM-Spin: A 0.5-to-1.2V Scalable Annealing Processor Using Digital Compute-In-Memory Spin Operators and Register-Based Spins for Combinatorial Optimization Problems},   year={2020},  volume={},  number={},  pages={480-482},  doi={10.1109/ISSCC19947.2020.9062938}}

@inproceedings{Centaur,
  title={Centaur: A chiplet-based, hybrid sparse-dense accelerator for personalized recommendations},
  author={Hwang, Ranggi and Kim, Taehun and Kwon, Youngeun and Rhu, Minsoo},
  booktitle={2020 ACM/IEEE 47th Annual International Symposium on Computer Architecture (ISCA)},
  pages={968--981},
  year={2020},
  organization={IEEE}
}


\end{document}